\documentclass[onecolumn,10pt]{IEEEtran}
\usepackage[utf8]{inputenc}
\renewcommand\baselinestretch{1.2}
\usepackage{amsmath} 
\usepackage{amsfonts,amsmath,amssymb,amsthm}
\usepackage{setspace,color}
\usepackage{amsbsy}
\usepackage{graphicx,multirow,bm}
\graphicspath{ {./figures/} }
\usepackage{color}
\usepackage{braket}
\usepackage{algorithm}
\usepackage{subcaption}
\usepackage{algpseudocode}
\usepackage{geometry} 
\allowdisplaybreaks[4]

\newtheorem{theorem}{Theorem}
\newtheorem{example}{Example}
\newtheorem{definition}{Definition}
\newtheorem{claim}{Claim}

\newtheorem{corollary}{Corollary}
\newtheorem{lemma}{Lemma}
\newtheorem{construction}{Construction}

\renewcommand{\c}{\boldsymbol{c}}
\newcommand{\GF}{\text{GF}}
\newcommand{\C}{\mathcal{C}}

\newcommand{\E}{\mathbb{E}}
\renewcommand{\d}{\boldsymbol{d}}

\DeclareMathOperator{\polylog}{polylog}
\DeclareMathOperator{\poly}{poly}

\title{
Secure Codes with List Decoding
\thanks{
Y. Gu is with the 
Faculty of Information Science and Electrical Engineering, Kyushu University, Fukuoka, Japan.
(email: gu@inf.kyushu-u.ac.jp)

I. Vorobyev is with Institute of Communications Engineering, 
Technical University of Munich, 
Munich, Germany. 
(email: ilya.vorobyev@tum.ru)

Y. Miao is with the Faculty of Engineering, Information and Systems, University of Tsukuba, Ibaraki, Japan.
(email: miao@sk.tsukuba.ac.jp)

This work was supported by RFBR and the National Science Foundation of Bulgaria (NSFB) under Project No. 20-51-18002, by BMBF-NEWCOM under Grant No. 16KIS1005, 
by JSPS under Grants No. 18H01133, No. 21K13830, and by Japan-Russia Research Cooperative Program between JSPS and RFBR under Project No. JPJSBP 120204802. 
An earlier version of this paper was presented in part at the IEEE International Symposium on Information Theory (ISIT2022) \cite{GVM2022}.
}
}

\author{\vskip 0.3cm 
Yujie Gu, Ilya Vorobyev, Ying Miao
}

\begin{document}
\maketitle

\begin{abstract}
In this paper we consider combinatorial secure codes in traitor tracing for protecting copyright of multimedia content. 
First, we introduce a new notion of secure codes with list decoding (SCLDs) for collusion-resistant multimedia fingerprinting, which includes many existing types of fingerprinting codes as special cases. 
Next, we build efficient identifying algorithms for SCLDs with complete traceability and establish bounds on its largest possible code rate. 
In comparison with the existing fingerprinting codes, it is shown that SCLDs have not only much more efficient traceability than separable codes but also a much larger code rate than frameproof codes. As a byproduct, new bounds on the largest code rate of binary separable codes are established as well. 
Furthermore, a two-stage dynamic traitor tracing framework is proposed for multimedia fingerprinting in the dynamic scenario, which could not only efficiently achieve the complete traceability but also provide a much larger capacity than the static scenario. 
\end{abstract}

\begin{IEEEkeywords}
Secure code, list decoding, dynamic traitor tracing, copyright protection, binary code, code rate 
\end{IEEEkeywords}

 \section{Introduction}
\label{sec:introduction}

The development and ubiquity of communication networks tremendously boost the spread and utility of multimedia content, such as text, audio, images, animations, and video, which, accordingly, stirs up the impending and challenging task of guaranteeing that the multimedia content is utilized for its intended purpose by authorized and legitimate consumers. 
For the sake of holding back multimedia content from being maliciously redistributed, 
digital fingerprinting has been proposed with the advantage that fingerprints can be embedded in multimedia content through watermarking techniques~\cite{Cox} and the malicious authorized consumers can be identified once they illegally use their content for unintended purpose~\cite{BS,Chor}.    

The orthogonal modulation and code modulation are two typical methods of embedding fingerprints into multimedia content~\cite{TWWL}. 
This work is considered with the latter code modulation scenario, which could accommodate more users than the former orthogonal modulation with the same amount of orthogonal signals~\cite{TWWL} and is briefly reviewed as follows. 
Suppose the host signal is a real vector $\mathbf{x}\in \mathbb{R}^m$ of length $m$. 
In code modulation, there are $n$ orthonormal basis signals $\{\mathbf{u}_i\in \mathbb{R}^m: 1\le i\le n\}$ which are typically not known to the users, and a watermark signal, or a \textit{fingerprint}, $\mathbf{w}_j$ is generated in the way that 
    $\mathbf{w}_j = \sum_{i=1}^n \mathbf{c}_j(i)\mathbf{u}_i$
where $\mathbf{c}_j(i)\in \{0,1\}$ according to the on-off keying (OOK) modulation. Accordingly user $j$ will be allocated a fingerprinted signal copy $\mathbf{y}_j=\mathbf{x}+\mathbf{w}_j$ of $\mathbf{x}$, which is feasible due to the fact that multimedia data is perceptually insensitive to minor perturbation in the data values~\cite{TWWL}.      
It is readily seen that there is a one-to-one correspondence between the authorized user $j$ and fingerprint $\mathbf{w}_j$, 
or equivalently the coefficient vector $\mathbf{c}_j=(\mathbf{c}_j(1),\ldots,\mathbf{c}_j(n))\in \{0,1\}^n$. 
Accordingly, all $M$ authorized users are associated with a collection of fingerprints $\{\mathbf{c}_j\in \{0,1\}^n: 1\le j\le M\}$. 
Regarding the collusion attacks, as pointed out in~\cite{TWWL}, since different bits of fingerprints that are additively embedded in multimedia may not be easily identifiable and arbitrarily manipulated, thereby linear collusion attacks such as averaging several fingerprinted signals are often more feasible for multimedia. 
In a \textit{linear attack}, a coalition consisting of a set of malicious authorized users $J\subseteq \{1,\ldots,M\}$ creates a forged copy $\hat{\mathbf{y}}$ by taking a linear combination of their copies $\mathbf{y}_j$, namely, 
\begin{align}\label{def:y=x+w}
    \hat{\mathbf{y}} = \sum_{j\in J} \lambda_j \mathbf{y}_j
    = \sum_{j\in J} \lambda_j  (\mathbf{x}+\mathbf{w}_j)
    = \mathbf{x}+ \sum_{j\in J} \lambda_j  \mathbf{w}_j
    = \mathbf{x}+ \sum_{i=1}^n \bigg(\sum_{j\in J} \lambda_j \mathbf{c}_j(i)\bigg) \mathbf{u}_i
\end{align}
where $\lambda_j$ are some real-valued coefficients such that 
$0<\lambda_j<1$ and $\sum_{j\in J} \lambda_j = 1$.  
In particular, if $\lambda_j = 1/|J|$ for all $j\in J$, it is called an \textit{averaging attack}.  
The \textit{traceability} refers to that once the forged copy $\hat{\mathbf{y}}$ is captured, partial or all of the colluders/traitors in $J$ would be identified. 
In particular, if all colluders in $J$ could be identified, it is referred to as the \textit{complete traceability}.      
In the identifying phase, the useful information can be extracted from the captured $\hat{\mathbf{y}}$ via the inner product $\langle\hat{\mathbf{y}}, \mathbf{u}_i\rangle  = \sum_{j\in J} \lambda_j \mathbf{c}_j(i)$ for all $1\le i\le n$. It is easily verified that $\langle\hat{\mathbf{y}}, \mathbf{u}_i\rangle\in [0,1]$ and 
\begin{itemize}
    \item $\langle\hat{\mathbf{y}}, \mathbf{u}_i\rangle = 0  \ 
   \text{implies}\ \{\mathbf{c}_j(i): j\in J\}=\{0\}$; 
   \item $\langle\hat{\mathbf{y}}, \mathbf{u}_i\rangle = 1  \ 
   \text{implies}\ \{\mathbf{c}_j(i): j\in J\}=\{1\}$; 
   \item $\langle\hat{\mathbf{y}}, \mathbf{u}_i\rangle \in (0,1) \ 
   \text{implies}\ \{\mathbf{c}_j(i): j\in J\}=\{0,1\}$. 
\end{itemize}
The yielded vector $\big(\{\mathbf{c}_j(1): j\in J\},\ldots,\{\mathbf{c}_j(n): j\in J\}\big)$ is termed as the \textit{evidence vector} 
or \textit{descendant code} of $J$ and will be used in the identifying/decoding algorithms, whose precise definition is referred to Section~\ref{sec:prelimina} as well. 
In terms of the anti-collusion fingerprinting, it is desirable to design a multimedia fingerprinting code which is a collection of fingerprints $\{\mathbf{c}_j: 1\le j\le M\}$ with efficient (partial or complete) traceability. 
This setting is typically referred to as the \textit{static} model of traitor tracing, with particular applications to electronic data distribution systems.

On the other hand, Fiat and Tassa~\cite{FT1999,FT2001} introduced the concept of \textit{dynamic} traitor tracing, which has numerous practical applications in protecting intellectual rights of \textit{streaming} data in broadband multicast systems. The dynamic traitor tracing typically allows to identify all the traitors in several stages; and in each stage, one could exploit the feedback from the previous and adapt the tracing strategies accordingly, see \cite{BPS2001,LDR2013,SaW2003} for example. Correspondingly, it allows the usage of several (different) secure/fingerprinting codes on the fly, which typically could accommodate much more users and/or provide more efficient traitor tracing in comparison with the static system.

In the literature, several classes of combinatorial multimedia fingerprinting codes have been proposed, which are briefly reviewed as follows. 
In 1998, Boneh and Shaw~\cite{BS} defined the $t$-frameproof codes, which could be utilized to trace back to all traitors in linear time $O(nM)$ where $n$ and $M$ are the length and the size of the code, respectively. Note that the definition of $t$-frameproof codes coincides with $(t, 1)$-separating codes, considered in~\cite{friedman1969universal,sagalovich1965method,sagalovich1994separaring} much earlier.
Later in order to neutralize the averaging attack in multimedia fingerprinting,
the AND anti-collusion code~\cite{TWWL}, 
the logical anti-collusion code (i.e. binary separable codes)~\cite{CM}, and strongly separable codes~\cite{JCM} have been introduced respectively, which all could guarantee the complete traceability and have been studied in e.g. \cite{FD2022,GFM2020,W2021,YZG2017}. 
Recently, signature codes with complete traceability for collusion-resistant multimedia fingerprinting have been investigated in~\cite{EFKL,FGHM}, which are shown to be essentially equivalent to uniquely decodable codebooks for weighted binary adder channel communication. In addition, signature codes with noise have been discussed in~\cite{EFKL,EFKM,FGHM,Voro} as well.

The list decoding technique has been applied to the traitor tracing problem (see e.g. \cite{BK2004,EFK2019,FMS2011,FS2004,SSW2001}), while the application model therein is related but different with this paper.   
In \cite{EFKL2016}, the authors constructed multimedia fingerprinting codes with efficient decoding but rather small code rate, which is based on the code concatenation together with the fast list decoding of Reed-Solomon codes with large distance. 
So far, all the existing applications of list decoding in traitor tracing are to construct certain types of fingerprinting codes by means of error-correcting codes with large Hamming distance and the concatenation construction, whereby the decoding analyses typically rely on the efficient list decoding algorithms of Reed-Solomon codes or algebraic-geometry codes (see e.g. \cite{GS1998}). 
In contrast to these known results, this work initially develops the list decoding property directly from the underlying traitor tracing model instead of taking error-correcting codes as an intermediary, and the list decoding here naturally goes beyond the Hamming metric according to the practical model.

In this paper, first, we introduce the notion of secure codes with list decoding (SCLDs) for collusion-resistant multimedia fingerprinting, which integrates the idea of list decoding into anti-collusion secure codes and hereby leads to efficient identifying algorithms. It is shown that SCLDs include many existing fingerprinting codes as special cases. 
Next, we build a two-step identifying algorithm for SCLDs and show that it could have the complete traceability as frameproof codes, which, however, is much more efficient than the traitor tracing based on the existing separable codes. 
On the other hand, we establish bounds on the code rates of binary and $q$-ary SCLDs respectively, which show that SCLDs could have a much larger code rate than the existing frameproof codes. As a byproduct, we provide new
lower bounds on the largest code rate of binary separable codes. 
In addition, some explicit constructions for SCLDs and efficient decoding algorithms for certain SCLDs with algebraic structures are provided as well. 
Furthermore, we discuss the dynamic multimedia fingerprinting and establish a two-stage dynamic traitor tracing framework based on the list decoding property, which provides not only efficient decoding but also much larger code rate (i.e. accommodate much more users) than the static scenario.

The paper is organized as follows. 
Section~\ref{sec:prelimina} presents the notations and definitions of codes. Section~\ref{sec:decoding of SCLD} exhibits the decoding algorithm for SCLDs. 
Section~\ref{sec: binary SCLD} establishes bounds on the code rates of binary SCLDs with large list size and binary separable codes. 
Section \ref{sec: binary SCLD constant list size} provides a lower bound on the largest code rate of binary SCLDs with constant list size.  
Section \ref{sec: q-ary SCLD} establishes bounds for $q$-ary SCLDs. 
Section \ref{sec:constructions} presents some explicit constructions of SCLDs and the corresponding efficient decoding algorithm. 
Section \ref{sec:dynamic_traitor_tracing} discusses the dynamic two-stage traitor tracing.   
Finally Section~\ref{sec:conclusion} concludes this paper.

\section{Preliminaries} 
\label{sec:prelimina}

To define the codes, we first introduce some notations. 
Let $n,q, M$ be positive integers and $Q\triangleq \{0,1,\ldots,q-1\}$ be an alphabet of size $|Q|=q$. 
For a prime power $q$, let $\text{GF}(q)$ denote the finite field of order $q$. 
Denote $[M]\triangleq \{1,2,\ldots,M\}$. 
A set of $M$ vectors $\mathcal{C}=\{\mathbf{c}_1,\mathbf{c}_2, \ldots, \mathbf{c}_M\}\subseteq Q^n$ is called an \textit{$(n, M, q)$ code} and each $\mathbf{c}_i=(\mathbf{c}_i(1),\mathbf{c}_i(2),\ldots,\mathbf{c}_i(n))\in Q^n$ is called a \textit{codeword} of length $n$. 
An $(n, M, q)$ code is also called a \textit{$q$-ary} code with \textit{length} $n$ and \textit{size} (or \textit{cardinality}) $M$. The \textit{rate} of an $(n,M,q)$ code $\mathcal{C}$ is $R(\mathcal{C}) = (\log_q M)/n$.

For code $\mathcal{C}\subseteq Q^n$, we define the $i$th projection 
of $\mathcal{C}$ as
\begin{equation*}
\mathcal{C}(i)=\{\mathbf{c}(i)\in Q: \mathbf{c}\in \mathcal{C}\},\ 1\le i\le n.
\end{equation*}
The \textit{descendant code} or \textit{evidence vector} of $\mathcal{C}$ is defined as
\begin{equation*}
\text{desc}(\mathcal{C})
=
(\mathcal{C}(1), \ldots,  \mathcal{C}(n))\in \mathcal{P}(Q)^n,
\end{equation*}
where $\mathcal{P}(Q)^n =  \mathcal{P}(Q)\times \cdots \times  \mathcal{P}(Q)$ and $\mathcal{P}(Q)$ is the \textit{power set} of $Q$,
i.e. $\mathcal{P}(Q) \triangleq  \{Q_0: Q_0\subseteq Q\}.$ 
For instance, 
if $\mathcal{C}=\{(0,1,0),(0,0,1),(0,1,2)\}$, then 
$\text{desc}(\mathcal{C})=(\{0\},\{0,1\},\{0,1,2\})$. 
In particular, let 
\begin{align*}
\text{Desc}_t(\mathcal{C})=
\{\text{desc}(\mathcal{C}_0) : \mathcal{C}_0\subseteq \mathcal{C}, 1\le |\mathcal{C}_0|\le t\}.
%
\end{align*}
%
For any $\mathbf{a}\in Q^n$ and $\mathbf{b}\in  \mathcal{P}(Q)^n$, we say that $\mathbf{a}$ is \textit{covered} by $\mathbf{b}$, denoted by $\mathbf{a}\preceq \mathbf{b}$, if and only if $\mathbf{a}(i)\in \mathbf{b}(i)$ for all $i\in [n]$.

\subsection{Secure code with list decoding}
\label{subsec:SCLD}

First we introduce the notion of secure codes with list decoding.  

\begin{definition}\label{def-scld}
Suppose that $\mathcal{C}=\{\mathbf{c}_1,\ldots,\mathbf{c}_M\}\subseteq Q^n$ is an $(n, M, q)$ code and $t\geq 2$, $t\le L\le M$ are  integers. 
Then $\mathcal{C}$ is a \textit{$\bar{t}$-secure code with list decoding}, or $\bar{t}$-SCLD$(n, M, q; L)$, if 
\begin{itemize}
    \item[(1)] for all distinct $\mathcal{C}_1,\mathcal{C}_2\subseteq \mathcal{C}$ with $|\mathcal{C}_1|\le t$ and $|\mathcal{C}_2|\le t$, we have $\text{desc}(\mathcal{C}_1)\ne \text{desc}(\mathcal{C}_2)$;
    \item[(2)] for any evidence vector $\mathbf{d}\in \text{Desc}_t(\mathcal{C})$, there are at most $L$ codewords covered by $\mathbf{d}$, namely 
    \begin{equation*}
        \text{Res}_{\mathcal{C}}(\mathbf{d})\triangleq|\{\mathbf{c}\in \mathcal{C}: \mathbf{c}\preceq \mathbf{d}\}|\le L.
    \end{equation*}
\end{itemize}
\end{definition}

We remark that the condition (1) of Definition~\ref{def-scld} guarantees the complete traceability 
of the secure code; the condition (2) indicates the list decoding of the secure code, which could induce efficient traitor tracing based on a two-step decoding (see our Algorithm~\ref{Alg1}). 
We illustrate this new concept with two small examples. 

\begin{example} It is readily checked that  
\begin{enumerate}
\item[1)]
$\mathcal{C}_1=\{(0,0,1),(1,0,1),(1,1,0)\}\subseteq \{0,1\}^3$ is 
a $\bar{2}$-SCLD$(3,3,2;3)$; and  
\item[2)] $\mathcal{C}_2=\{(0,0,0),(1,0,0),(0,1,0),(0,0,1)\}\subseteq \{0,1\}^3$ is a $\bar{2}$-SCLD$(3,4,2;3)$.
\end{enumerate}
Note that in $\mathcal{C}_1$ the list size $L = M = t+1=3$, while in $\mathcal{C}_2$ the list size $L=M-1=t+1=3$. 
\end{example}

It is worth noting that SCLDs can be seen as a unified concept of multimedia fingerprinting codes with complete traceability in terms of the list size. Recall three classes of existing codes from~\cite{BS,CM,shchukin2016list}. 
\begin{definition}
Suppose that $\mathcal{C}=\{\mathbf{c}_1,\ldots,\mathbf{c}_M\}\subseteq Q^n$ is an $(n, M, q)$ code and $t\geq 2$ is an integer. Then 
\begin{enumerate}
    \item[1)] $\mathcal{C}$ is a \textit{$t$-frameproof code}, or $t$-FPC$(n, M, q)$, if for any $\mathcal{C}_0\subseteq \mathcal{C}$ such that $|\mathcal{C}_0|=t$ and any $\mathbf{a}\in \mathcal{C}\setminus \mathcal{C}_0$, it holds that $\mathbf{a}\npreceq \text{desc}(\mathcal{C}_0)$; 
    \item[2)] $\mathcal{C}$ is a \textit{$\bar{t}$-separable code}, or $\bar{t}$-SC$(n, M, q)$, if 
    for all distinct $\mathcal{C}_1,\mathcal{C}_2\subseteq \mathcal{C}$ with $|\mathcal{C}_1|\le t$ and $|\mathcal{C}_2|\le t$, we have $\text{desc}(\mathcal{C}_1)\ne \text{desc}(\mathcal{C}_2)$. 
    \item[3)] $\mathcal{C}$ is a \textit{$t$-hypercode with list decoding}, or $t$-HLD$(n, M, q;$ $L)$, if for any $\mathcal{C}_0\subseteq \mathcal{C}$ such that $|\mathcal{C}_0|=t$, and $\mathbf{d}= \text{desc}(\mathcal{C}_0)$, we also have 
\vspace*{-1mm}
\begin{equation*}
    \text{Res}_{\mathcal{C}}(\mathbf{d})=|\{\mathbf{c}\in \mathcal{C}: \mathbf{c}\preceq \mathbf{d}\}|\le L.
\end{equation*}
\end{enumerate}
\end{definition}

It is readily verified from the above definitions that a $t$-HLD$(n,M,q;L)$ implies a $t'$-HLD$(n,M,q;L)$ for any $t'\le t$, 
and $t$-FPCs and $\bar{t}$-SCs are in fact special cases of $\bar{t}$-SCLDs. 
More precisely, we have the followings immediately. 
\begin{lemma}\label{lemma:code-relations}
Let $t,n,q,M,L$ be positive integers such that $M\ge L\ge t$, then the followings hold. 
\begin{enumerate}
    \item[1)] 
    A $t$-FPC$(n, M, q)$ is equivalent to a $\bar{t}$-SCLD$(n, M, q; t)$. 
    \item[2)] 
    A $\bar{t}$-SC$(n, M, q)$ is equivalent to a $\bar{t}$-SCLD$(n, M, q; M)$. 
\item[3)] 
    Let $\mathcal{C}$ be an $(n,M,q)$ code. Then $\mathcal{C}$ is a
     $\bar{t}$-SCLD$(n, M, q;L)$ if and only if $\mathcal{C}$ is a $\bar{t}$-SC$(n, M, q)$ and a $t$-HLD$(n, M, q; L)$ simultaneously.  
  \item[4)] 
    A $\bar{t}$-SCLD$(n, M, q;L)$ is a $\bar{t}$-SCLD$(n, M, q;L')$ for any $L\le L'\le M$.
\end{enumerate}
\end{lemma}

\subsection{Code rate}

Let $M_{\rm FPC}(t,n,q)$, $M_{\rm SCLD}(\bar{t},n,q;L)$, $M_{\rm HLD}(t,n,q;L)$, 
$M_{\rm SC}(\bar{t},n,q)$ denote the largest cardinality of a $q$-ary $t$-FPC, $\bar{t}$-SCLD, $t$-HLD,  $\bar{t}$-SC of length $n$,  respectively. 

According to practical applications~\cite{TWWL}, binary fingerprinting codes are typically desired. 
It is well known that, by Forney concatenation~\cite{Forney}, binary codes can be derived from general $q$-ary codes as well.  
Hence in what follows we will consider the following two typical scenarios. 
\begin{itemize}
\item For binary codes, 
denote their largest asymptotic code rates as
\begin{align*}\label{def-limits}
R_{\rm FPC}(t)&=\limsup_{n\to\infty}\frac{\log_2 M_{\rm FPC}(t,n,2)}{n}, \\
R_{\rm SCLD}(\bar{t};L)&=\limsup_{n\to\infty}\frac{\log_2 M_{\rm SCLD}(\bar{t},n,2;L)}{n},\\ 
R_{\rm HLD}(t;L)&=\limsup_{n\to\infty}\frac{\log_2 M_{\rm HLD}(t,n,2;L)}{n},\\ 
R_{\rm SC}(\bar{t})&=\limsup_{n\to\infty}\frac{\log_2 M_{\rm SC}(\bar{t},n,2)}{n}. 
\end{align*}

\item For $q$-ary codes of length $n$, 
denote their largest asymptotic code rates as
\begin{align*}
R_{\rm FPC}(t,n)&=\limsup_{q\to\infty}\frac{\log_q M_{\rm FPC}(t,n,q)}{n}, \\
R_{\rm SCLD}(\bar{t},n;L)&=\limsup_{q\to\infty}\frac{\log_q M_{\rm SCLD}(\bar{t},n,q;L)}{n},\\ 
R_{\rm HLD}(t,n;L)&=\limsup_{q\to\infty}\frac{\log_q M_{\rm HLD}(t,n,q;L)}{n},\\ 
R_{\rm SC}(\bar{t},n)&=\limsup_{q\to\infty}\frac{\log_q M_{\rm SC}(\bar{t},n,q)}{n}. 
\end{align*} 
\end{itemize}

According to Lemma~\ref{lemma:code-relations}, we immediately have the following consequences. 
\begin{corollary}
Let $t,n,L,L'$ be integers such that $2 \le t\le L\le L'\le  M$, where $M$ is the cardinality of the code. Then we have 
\begin{itemize}
    \item[1)] $R_{\rm FPC}(t)\le R_{\rm SCLD}(\bar{t};L) \le \min\{R_{\rm SC}(\bar{t}),R_{\rm HLD}(t;L)\}$.
    \item[2)] $R_{\rm FPC}(t,n)\le R_{\rm SCLD}(\bar{t},n;L) \le R_{\rm SC}(\bar{t},n)$. 
    \item[3)] $R_{\rm SCLD}(\bar{t},n;L) \le R_{\rm SCLD}(\bar{t},n;L')$. 
\end{itemize}
\end{corollary}

We summarize the state-of-the-art bounds for FPCs, HLDs, SCs and the new bounds for SCLDs in the case when the alphabet size $q$ is sufficiently large in Table \ref{table:existing_bounds}.

\begin{table}[h]\centering
\caption{\ \ Asymptotic code rates of $q$-ary FPCs, SCLDs, HLDs, SCs as $q\to \infty$}
\label{table:existing_bounds}
\renewcommand{\arraystretch}{1.2}
\begin{tabular}{|c|c|c|c|c|}
\hline 
& $R_{\rm FPC}(t,n)$ & $R_{\rm SCLD}(\bar{t},n;L)$ & $R_{\rm HLD}(t,n;L)$ & $R_{\rm SC}(\bar{t},n)$ 
\\ 
\hline
\multirow{2}{*}{Code Rate} & 
\multirow{2}{*}{$=\frac{\lceil n/t \rceil}{n}$}   & 
$\ge \begin{cases} 2/3 & \text{if $t=2$, $L\ge 3$}
\\ 1/(t-1) & \text{if $t>2$, $L\ge t+1$}\end{cases}$ & \multicolumn{1}{|l|}{$\ge  \frac{L-t+1}{L}$} &
\multicolumn{1}{|l|}{$\ge \begin{cases} 2/3 & \text{if $t=2$}\\ 1/(t-1) & \text{if $t>2$}\end{cases}$} \\ 
\cline{3-5} 
 &  & \multicolumn{1}{|l|}{$\le \begin{cases} \lceil 2n/3\rceil /n & \text{if $t=2$}\\ \lceil n/(t-1)\rceil /n & \text{if $t>2$}\end{cases}$} & 
 $\le\frac{(L-t+1)\lceil(n/L\rceil}{n}$ & $\le \begin{cases} \lceil 2n/3\rceil /n & \text{if $t=2$}
\\ \lceil n/(t-1)\rceil /n & \text{if $t>2$}\end{cases}$ \\ \hline
Reference  & \cite{B03,cohen2003asymptotic}  &  Theorem~\ref{thm-lower-qnary}, \cite{B15,GG}  & \cite{d2019separable,shchukin2016list} &  \cite{B15,GG}\\ \hline
\end{tabular}
\end{table}

\section{Identifying algorithm for SCLDs}
\label{sec:decoding of SCLD}

In this section we present a two-step identifying algorithm for SCLDs and discuss its performance in comparison with the existing FPCs and SCs. A similar idea was used in the context of union-free codes 
for non-adaptive group testing as well~\cite{vorobyev2021fast}.

\begin{theorem}\label{thm:decoding}
A $\bar{t}$-SCLD$(n,M,q;L)$ has an identifying  
algorithm with complete traceability and $O\big(\max\{nM,nL^t\}\big)$ time complexity. 
\end{theorem}
\begin{IEEEproof}
The identifying algorithm for a $\bar{t}$-SCLD$(n,M,q;L)$ can be done as Algorithm~\ref{Alg:SCLD}, where 
\begin{itemize}
    \item the time cost of Step 1 is $O(nM)$ and $|\mathcal{W}|\le L$; 
    \item the time cost of Step 2 is $O(n|\mathcal{W}|^t)=O(nL^t).$
\end{itemize}

\begin{algorithm}
\caption{Identifying algorithm for $\bar{t}$-SCLDs}\label{Alg:SCLD}\label{Alg1}
\textbf{Input:} $\bar{t}$-SCLD code $\mathcal{C}$; the evidence vector $\mathbf{d}\in \text{Desc}_t(\mathcal{C})$\\
\textbf{Output:} the set of all traitors $\mathcal{T}$
\begin{algorithmic}[1]
\State $\mathcal{T} = \emptyset $, $\mathcal{W} = \emptyset $. \Comment{Initialize the candidate sets}
\For{each $\mathbf{c} \in \mathcal{C}$} 
\Comment{Step 1}
\If {$\mathbf{c}\preceq \mathbf{d}$}
\State $\mathcal{W} = \mathcal{W}\cup \{\mathbf{c}\} $;
\EndIf
\EndFor
\For{each subset $\mathcal{S}\subseteq \mathcal{W}$ of size at most $t$} 
\Comment{Step 2}
\If {$\text{desc}(\mathcal{S}) == \mathbf{d}$}
\State $\mathcal{T} = \mathcal{S}$;
\State output $\mathcal{T}$; 
\EndIf
\EndFor
\end{algorithmic}
\end{algorithm} 

Therefore the total time cost of Algorithm~\ref{Alg:SCLD} is $O(\max\{nM,nL^t\})$. 
Furthermore, according to Definition~\ref{def-scld} of SCLDs, the true coalition set $\mathcal{T}$, after Step 1, is a subset of $\mathcal{W}$, and could be exactly identified after Step 2. That is, Algorithm~\ref{Alg:SCLD} for SCLDs is with complete traceability, as desired. 
\end{IEEEproof}

By taking $L = \lfloor M^{1/t} \rfloor $ in Theorem~\ref{thm:decoding}, we have an immediate corollary. 
\begin{corollary}\label{coro:SCLD-decoding}
A $\bar{t}$-SCLD$\big(n,M,q;\lfloor M^{1/t} \rfloor\big)$ has an identifying algorithm with complete traceability and $O(nM)$ time complexity. 
\end{corollary}

Recall the traitor tracing of FPCs and SCs from~\cite{CM}.  
\begin{lemma}[\cite{CM}]\label{lemma:recall-decoding}
\begin{enumerate}
    \item[1)] A $t$-FPC$(n,M,q)$ has an identifying  
algorithm with complete traceability and $O(nM)$ time complexity. 
    \item[2)] A $\bar{t}$-SC$(n,M,q)$ has an identifying  
algorithm with complete traceability and $O(nM^t)$ time complexity. 
\end{enumerate}
\end{lemma}

It is readily seen from Corollary~\ref{coro:SCLD-decoding} and Lemma~\ref{lemma:recall-decoding} that the newly proposed SCLDs could have the same traceability as FPCs, which is much more efficient than that of SCs. In the next section, we will show that SCLDs could have a much larger code rate than FPCs.

\section{Bounds for binary $\bar{t}$-SCLDs}
\label{sec: binary SCLD}
In this section we establish bounds on the largest asymptotic code rate of binary SCs, HLDs, and SCLDs, respectively.  


\subsection{Lower bounds for binary SCs, SCLDs, HLDs} 

According to Theorem~\ref{thm:decoding}, it is desirable to consider the case of list size $L=M^{\alpha}$ with $\alpha \in [0,1]$. To simplify the notation, we use $M^{(\alpha)}_{\rm SCLD}(\bar{t},n,q)$ and $M^{(\alpha)}_{\rm HLD}(t,n,q)$ 
to denote the largest cardinality of a $q$-ary $\bar{t}$-SCLD and a $t$-HLD with length $n$ and list size $L=M^{\alpha}$, respectively, where $M$ is the corresponding code size. 
Accordingly we denote their largest asymptotic binary code rates as
\begin{align}
R^{(\alpha)}_{\rm SCLD}(\bar{t})&=\limsup_{n\to\infty}\frac{\log_2 M^{(\alpha)}_{\rm SCLD}(\bar{t},n,2)}{n},\\ 
\label{eq:def-R-hld-alpha}
R^{(\alpha)}_{\rm HLD}(t)&=\limsup_{n\to\infty}\frac{\log_2 M^{(\alpha)}_{\rm HLD}(t,n,2)}{n}. 
\end{align}

Now we establish the following lower bounds for SCs, HLDs, and SCLDs using random coding with expurgation. Let $h(x)\triangleq -x\log_2(x)-(1-x)\log_2(1-x)$  be the binary entropy function. 

\begin{theorem}\label{thm-lower-scld}
\begin{enumerate}
\item[1)] $R_{\rm SC}(\bar{t})\ge \max\limits_{p\in (0,1)}\underline{R}_{\rm SC}(t, p)$,
    where
    \begin{align}
    \label{eq:rate of sep codes}
        &\underline{R}_{\rm SC}(t, p)
        \triangleq \min\limits_{\substack{1\le t_1\le t_2\le t,\\ 0\le m\le t_1,\\ m\ne t_2}}\left(\frac{-\log_2(1-P_{g}(t_1, t_2, m))}{t_1+t_2-m-1}\right),
    \end{align}
    and 
    \begin{align*}
	&P_{g}(t_1, t_2, m)\triangleq p^{t_1}+p^{t_2} + (1-p)^{t_1} + (1-p)^{t_2} 
	- 2p^{t_1+t_2-m} - 2(1-p)^{t_1+t_2-m} \ \ \ \text{for}\  m>0, \\
    &P_{g}(t_1, t_2, 0)\triangleq 1-
	p^{t_1+t_2}-(1-p)^{t_1+t_2}
	    -(1-p^{t_1}-(1-p)^{t_1})(1-p^{t_2}-(1-p)^{t_2}).
    \end{align*}
\item[2)]  For any $\alpha\in (0,1)$, we have  $R^{(\alpha)}_{\rm HLD}(t)\ge \max\limits_{p\in (0,1)}\underline{R}^{(\alpha)}_{\rm HLD}(t, p)$,
    where
    \begin{equation}\label{eq:rate of hld codes}
    \underline{R}^{(\alpha)}_{\rm HLD}(t, p)\triangleq \frac{h(p)-tph(1/t)}{1-\alpha}.
    \end{equation}
\item[3)] For any $\alpha\in (0,1)$, we have  $R^{(\alpha)}_{\rm SCLD}(\bar{t})\ge \max\limits_{p\in (0,1)}\min\{\underline{R}_{\rm SC}(t, p), \underline{R}^{(\alpha)}_{\rm HLD}(t, p)\}$.
\end{enumerate}
\end{theorem}

\begin{IEEEproof}
Let $\mathcal{C}=\{\c_1,\c_2,\ldots,\c_M\}$ be a collection of $M$ binary vectors of length $n$, in which each coordinate $\c_i(j)$ is chosen from $\{0,1\}$ independently at random and equals $1$ with probability $p$, where $p\in (0,1)$. 
This random ensemble will be used in the following arguments for all three claims.
	
1) Consider the requirements of a $\bar{t}$-SC. 
	A pair of distinct index sets $I_1\subseteq [M]$, $I_2\subseteq [M]$, $|I_1|=t_1$, $|I_2|=t_2$, $|I_1\cap I_2|=m$, $1 \le t_1, t_2\le t$, is called a \textit{bad $(t_1, t_2, m)$-pair} if $\text{desc}(\{\c_i: i \in I_1\})=\text{desc}(\{\c_i: i \in I_2\})$. 
	For notation simplicity, we denote 
	$\C_{I}=\{\c_i: i \in I\}\subseteq \C$ for any $I\subseteq [M]$.

	Next we estimate the expectation $E_1$ of the number of bad $(t_1, t_2, m)$-pairs for all $(t_1, t_2, m)\in \Xi$ where 
    \begin{align}\label{def-Xi}
     \Xi \triangleq  \{(t_1, t_2, m):  1\le t_1\le t_2\le t, 0\le m\le t_1, m\ne t_2\}.
    \end{align}
	To that end, denote 
	\begin{align*}
	    \mathcal{B}(t_1,t_2,m) \triangleq 
	    \{(I_1,I_2):\  
	    &I_1\subseteq [M], I_2\subseteq [M], |I_1|=t_1, |I_2|=t_2, |I_1\cap I_2|=m
	    \}.
	\end{align*}
	Clearly, $|\mathcal{B}(t_1,t_2,m)|\le M^{t_1+t_2-m}$. 
	Let $P_{g}(t_1, t_2, m)$ denote 
	the probability that $\C_{I_1}(i)$ and $\C_{I_2}(i)$ are different for an arbitrarily row $i$, where $(I_1,I_2)\in \mathcal{B}(t_1,t_2,m)$. 
	Then for any $(I_1,I_2)\in \mathcal{B}(t_1,t_2,m)$, the probability of that $(I_1,I_2)$ is bad is $(1-P_{g}(t_1, t_2, m))^n$. Now we compute $P_{g}(t_1, t_2, m)$ by discussing the following two cases. 
	
	\textbf{Case 1. }
	Consider the case of that the intersection $I_1\cap I_2$ is not empty, i.e. $m>0$. If $\C_{I_1}(i)\ne \C_{I_2}(i)$, then there are four possible options:
	\begin{itemize}
	    \item 
		$\C_{I_1}(i)=\{1\}$, $\C_{I_2}(i)=\{0, 1\}$;
		\item
		$\C_{I_1}(i)=\{0\}$, $\C_{I_2}(i)=\{0, 1\}$;
		\item
		$\C_{I_1}(i)=\{0, 1\}$, $\C_{I_2}(i)=\{1\}$;
		\item
		$\C_{I_1}(i)=\{0, 1\}$, $\C_{I_2}(i)=\{0\}$.
	\end{itemize}
The corresponding probabilities for these four cases are:
 \begin{itemize}
     \item 
 	$p^{t_1}(1-p^{t_2-m})$; 
 	\item
 	$(1-p)^{t_1}(1-(1-p)^{t_2-m})$; 
 	\item
 	$p^{t_2}(1-p^{t_1-m})$; 
 	\item
 	$(1-p)^{t_2}(1-(1-p)^{t_1-m})$.
 \end{itemize}
The total probability $P_{g}(t_1, t_2, m)$ for $m>0$ is the sum of these probabilities
	\begin{align*}
	P_{g}(t_1, t_2, m)&=p^{t_1}+p^{t_2} + (1-p)^{t_1} + (1-p)^{t_2} 
	- 2p^{t_1+t_2-m} - 2(1-p)^{t_1+t_2-m}.
	\end{align*}
	
	\textbf{Case 2.} Now consider the case of $I_1\cap I_2=\emptyset$, i.e. $m=0$. 
	We have 
	$$P_{g}(t_1, t_2, 0) =1-P_{\rm bad}(t_1, t_2)$$
	where $P_{\rm bad}(t_1, t_2)$ is the probability of $\C_{I_1}(i)$ and $\C_{I_2}(i)$ coincide in coordinate $i$. 
	Similar to Case 1, we obtain
	\begin{align*}
	    P_{\rm bad}(t_1, t_2)
	    &=p^{t_1+t_2}+(1-p)^{t_1+t_2}
	    +(1-p^{t_1}-(1-p)^{t_1})(1-p^{t_2}-(1-p)^{t_2}).
	\end{align*}
	
Based on the above, the expectation $E_1$ is 
	\begin{align*}
		E_1&=\sum\limits_{(t_1,t_2,m)\in \Xi} 
		|\mathcal{B}(t_1,t_2,m)|(1-P_{g}(t_1, t_2, m))^n\\
		&\le \sum\limits_{(t_1,t_2,m)\in \Xi} 
		 M^{t_1+t_2-m}(1-P_{g}(t_1, t_2, m))^n. 
	\end{align*}
Next we choose parameters $p\in (0,1)$ and $M=2^{nR}$ in such way that $E_1< cM/(n+1)$ for an arbitrary constant $c>0$ and 
sufficiently large $n$. Actually this can be satisfied if 
$
R<\underline{R}_{\rm SC}(t, p)
$
for $p\in (0,1)$,
where 
$\underline{R}_{\rm SC}(t, p)$ is defined in~\eqref{eq:rate of sep codes}. Take $c=1/4$. Then we remove one element from each bad pair.  
Note that there are no repeated vectors left since their corresponding indices have been removed as bad $(1,1,0)$-pairs. Then we can conclude that the obtained code is indeed a $\bar{t}$-SC with size greater than $M/2$ and the rate at least $\log_2(M/2)/n=R-o(1)$ as $n\to \infty$. 

2) Now we consider for a HLD. 
Recall the randomly generated $\mathcal{C}$.  
\begin{itemize}
    \item We call a vector $\c_i $ \textit{bad} if its weight is not equal to $\lfloor p(n+1)\rfloor$; otherwise call it \textit{good}. 
    Denote the set of all good vectors $\c_i\in \C$ as $\C_0$. 
    Let $E_2$ denote the mathematical expectation of the number of bad vectors in $\C$.
    \item 
    For $\C_0'\subseteq \C_0$ such that $|\C_0'|=t$, the subset $\C_0'$ is called a \textit{bad set} if the size of $\text{Res}_{\C_0}(\text{desc}(\C_0'))$ is greater than $L'$, where $L'=(\frac{M}{2n+2})^{\alpha}$. 
    Let $E_3$ denote the mathematical expectation of the number of bad sets in $\C_0$.
\end{itemize}
    
	Next we estimate $E_2$ and $E_3$ as follows. 
	
	(i) Consider $E_2$. 
	The weight of a vector 
	$\c\in \C$
	 can be seen as a random variable following the Binomial distribution with parameter $n$ and $p$, i.e. $B(n,p)$. 
	It is known that the \textit{mode} (i.e. the most frequent value) of $B(n,p)$ is $\lfloor p(n+1)\rfloor$. 
	Since there are totally $n+1$ possible weights,  the probability that a vector $\c$ is good (i.e. with weight $\lfloor p(n+1)\rfloor$) is no less than the average $1/(n+1)$. 
	Hence, the mathematical expectation $E_2$ of the number of bad vectors is at most $Mn/(n+1)$.
	
	(ii)
Consider $E_3$. First we compute the probability that one good vector $\c\in\C_0$ is covered by a fixed evidence vector $\d=\text{desc}(\C'_0)$, where $\C'_0\subseteq \C_0$, $|\C'_0|=t$ and $\c\notin \C'_0$.
Note that the number of coordinates $i$ such that $1\in \d(i)$ is at most $t\lfloor p(n+1)\rfloor$. Therefore, the probability that $\c \preceq \d$ can be upper bounded as 

$$
\text{Pr}(\c \preceq \d)\leq \frac{\binom{t\lfloor p(n+1)\rfloor}{\lfloor p(n+1)\rfloor}}{\binom{n}{\lfloor p(n+1)\rfloor}}=2^{n(tph(1/t)-h(p)+o(1))}  
$$
as $n\to \infty$. 
If there exist more than $L'=(\frac{M}{2n+2})^{\alpha}$ vectors in $\C_0$ that are covered by $\d$, then there exists a set of  exactly $L'-t$ vectors in $\C_0\setminus \C_0'$ that are covered by $\d$. Therefore, 
the probability that $\d$ covers more than
$L'$ 
vectors in $\C_0$ is less than
\begin{align*}
\binom{|\C_0\setminus \C_0'|} {L'-t}2^{n(tph(1/t)-h(p)+o(1))(L'-t)}
&\le
\binom{M}{L'-t}2^{n(tph(1/t)-h(p)+o(1))(L'-t)}\\
&\le
\frac{M^{L'-t}}{(L'-t)!}\left(M^{R^{-1}(tph(1/t)-h(p)+o(1))}\right)^{L'}\\
&\le
\frac{M^{L'-t}}{(L')^{-t}(L')!}\left(M^{R^{-1}(tph(1/t)-h(p)+o(1))}\right)^{L'}\\
&\le\frac{e^{L'}M^{L'-t}}{(L')^{L'-t}}\left(M^{R^{-1}(tph(1/t)-h(p)+o(1))}\right)^{L'}\\
&=M^{L'(1-\log_ML'+R^{-1}(tph(1/t)-h(p)+o(1)))}.
\end{align*}
Take $R<\underline{R}^{(\alpha)}_{\rm HLD}(t, p)$, where $\underline{R}^{(\alpha)}_{\rm HLD}(t, p)$ is defined in~\eqref{eq:rate of hld codes}.
For any such $R$ the probability that $\d$ covers more than $L'$ vectors in $\C_0$ is at most
$
M^{L'(-\varepsilon+o(1))}
$
for some positive $\varepsilon$.

Then the mathematical expectation $E_3$ is at most
$$
E_3<\binom{|\C_0|} {t}M^{L'(-\varepsilon+o(1))} < M/(4n+4) 
$$
for large enough $n$. 

Delete all bad vectors and one vector from each bad set. The obtained code is a $t$-HLD with list size $L'$ and cardinality $\ge \frac{M}{2n+2}$. Delete some additional vectors to obtain a code of size exactly $M'=\frac{M}{2n+2}$. The final code is a $t$-HLD with list size $L'=(M')^{\alpha}$ and rate $\frac{1}{n}\log_2\frac{M}{2n+2}=R-o(1)$ as $n\to \infty$.

3) Now we consider the requirements for SCLD based on the previous discussions on SC and HLD. 
Recall the same random ensemble $\C$. According to the arguments of the previous two claims, if we take $$R\le \min\{\underline{R}_{\rm SC}(t, p), \underline{R}^{(\alpha)}_{\rm HLD}(t, p)\}$$ 
then it holds that 
$$
E_1<M/(4n+4);\;
E_2\le Mn/(n+1);\;
E_3<M/(4n+4),
$$
which implies  
$
E_1+E_2+E_3<M-M/(2n+2).
$

Remove all bad vectors, one vector from each bad pair, and one vector from each bad set. The obtained code is a $\bar{t}$-SC and simultaneously a $t$-HLD, which, according to Lemma~\ref{lemma:code-relations}, is in fact a $\bar{t}$-SCLD, with code size $M'=M/(2n+2)$ and list size $L'=(M')^{\alpha}$. The code rate is at least $n^{-1}\log_2\frac{M}{2n+2}=R-o(1)$ as $n\to \infty$. The theorem is proved.
\end{IEEEproof}

\vskip 0.1cm
Based on the above Theorem~\ref{thm-lower-scld}, we alternatively have the following lower bound for HLDs. 

\begin{theorem}\label{thm:lower_hld}
Let $t\ge 2$ be a positive integer and $0< \alpha< 1$ be a constant. Then 
    \begin{equation}\label{eq_thm:rate of hld codes}
    R^{(\alpha)}_{\rm HLD}(t)\ge  \frac{1}{1-\alpha}\bigg[h\bigg(\frac{1}{2^{th(1/t)}+1}\bigg)-\frac{th(1/t)}{2^{th(1/t)}+1}\bigg].
    \end{equation}
Furthermore, if $t$ is sufficiently large, we have 
\begin{align}
     R^{(\alpha)}_{\rm HLD}(t)\ge
    \frac{\log_2 e}{et(1-\alpha)}(1+o(1)) \approx \frac{0.530738}{t(1-\alpha)}(1+o(1)),\quad (t\to \infty). 
\end{align}
\end{theorem}

\begin{IEEEproof}
Recall from Theorem~\ref{thm-lower-scld} that
    \begin{equation}\label{eq_recall:rate of hld codes}
    R^{(\alpha)}_{\rm HLD}(t)\ge \max\limits_{p\in (0,1)} \frac{h(p)-tph(1/t)}{1-\alpha}.
    \end{equation}
Let $F(p,t)\triangleq h(p)-tph(1/t)$. Then 
\begin{align*}
    \frac{\partial F(p,t)}{\partial p}
    =\log_2\bigg(\frac{1-p}{p}\bigg) - th(1/t). 
\end{align*}
Setting $\frac{\partial F(p,t)}{\partial p} = 0$ gives $p^*= \frac{1}{2^{th(1/t)}+1}$. 
Clearly $0 < p^* < 1$.
Then plugging $p=p^*$ into \eqref{eq_recall:rate of hld codes} yields the bound 
\eqref{eq_thm:rate of hld codes}.

Next we consider the case when $t$ is sufficiently large. 
Taking $p=\frac{1}{et}$, where $e\approx 2.71828$ is the Euler's number, into $F(p,t)$ yields 
\begin{align*}
    F\bigg(p=\frac{1}{et},t\bigg) 
    &= 
    h\bigg(\frac{1}{et}\bigg) - \frac{1}{e}h\bigg(\frac{1}{t}\bigg)\\
    &=
    -\frac{1}{et}\log_2\bigg(\frac{1}{et}\bigg)
    -\bigg(1-\frac{1}{et}\bigg)\log_2\bigg(1-\frac{1}{et}\bigg) + \frac{1}{et}\log_2\bigg(\frac{1}{t}\bigg)+\frac{1}{e}\bigg(1-\frac{1}{t}\bigg)\log_2\bigg(1-\frac{1}{t}\bigg) \\
    &=
    \frac{1}{et}\log_2e
    -\bigg(1-\frac{1}{et}\bigg)\log_2\bigg(1-\frac{1}{et}\bigg) + \frac{1}{e}\bigg(1-\frac{1}{t}\bigg)\log_2\bigg(1-\frac{1}{t}\bigg) \\
    &= 
    \frac{1}{et}\log_2e
    +\log_2e\cdot\bigg(1-\frac{1}{et}\bigg)
    \bigg[\frac{1}{et} + \frac{1}{2(et)^2}+\cdots \bigg]- \frac{\log_2e}{e}\cdot\bigg(1-\frac{1}{t}\bigg)
    \bigg[\frac{1}{t} + \frac{1}{2t^2}+\cdots \bigg]\\
    &=\frac{1}{et}\log_2e 
    + \frac{\log_2e}{et} - \frac{\log_2e}{e}\cdot \frac{1}{t} +o\bigg(\frac{1}{t}\bigg)\\
    & = \frac{1}{et}\log_2e \cdot(1+o(1)),
\end{align*}
where the fourth equality follows from the Taylor series that $\ln(1-x) = -\sum^{\infty}_{n=1} \frac{x^n}{n}= - x - \frac{x^2}{2} -\cdots $ for any  $x\in [-1,1]$. 
Hence we conclude that 
\begin{align*}
    R^{(\alpha)}_{\rm HLD}(t)\ge \max\limits_{p\in (0,1)} \frac{F(p,t)}{1-\alpha}
    \ge 
    \frac{F(p=\frac{1}{et},t)}{1-\alpha}
    = \frac{\log_2e}{et(1-\alpha)}\cdot(1+o(1))
    \approx 
    \frac{0.530738}{t(1-\alpha)}(1+o(1)), \quad (t\to \infty)
\end{align*}
as desired. 
\end{IEEEproof}

It is worth noting that the case when $\alpha\to 0$ 
and $t$ goes to infinity implies that the list size $L$ also goes to infinity, whereas the list size $L\ll M$ where $M$ is the corresponding code cardinality.  
For this particular case, the above Theorem~\ref{thm:lower_hld} indicates that $\lim_{\alpha\to 0}R^{(\alpha)}_{\rm HLD}(t)\ge \frac{0.530738}{t}(1+o(1))$, which coincides with the bound derived in \cite[Theorem 3]{shchukin2016list}. 

\subsection{Remarks}
Table~\ref{table:numerical_results} illustrates numerical values of the state-of-the-art lower bounds on the largest code rates of binary SCs, FPCs, SCLDs, in which the values in bold are derived from Theorem~\ref{thm-lower-scld};     
$R_{\rm SC}(\bar{2})$ is from \cite{L69}; $R_{\rm FPC}(2)$ is from \cite{R2013}; and $R_{\rm FPC}(t)$ for $t>2$ is from~\cite{shchukin2016list}.

\begin{table}[h]\centering
	\caption{Lower bounds from Theorem~\ref{thm-lower-scld} and \cite{L69,R2013,shchukin2016list}}
	\label{table:numerical_results}
\renewcommand{\arraystretch}{1.2}
	\begin{tabular}{|c|c|c|c|c|c|c|c|c|c|}
		\hline
		$t$             & 2       & 3       & 4       & 5       & 6 \\ \hline
		$R_{\rm SC}(\bar{t})\ge $   & 0.5     & \textbf{0.13834} & \textbf{0.06198} & \textbf{0.03138} & \textbf{0.02003}  \\ \hline
		$R^{(1/t)}_{\rm SCLD}(\bar{t})\ge $ & \textbf{0.44452} & \textbf{0.13205} & \textbf{0.05770} & \textbf{0.03105} & \textbf{0.01997} \\ \hline
		$R_{\rm FPC}(t)\ge $   & 0.20756  & 0.07999  & 0.04392  & 0.02794  & 0.01936 \\ \hline
	\end{tabular}
\end{table}

It is worth noting that the bounds  for $R^{(1/t)}_{\rm SCLD}(\bar{t})$ and $R_{\rm SC}(\bar{t})$ are quite close, and much larger than $R_{\rm FPC}(t)$ (in particular, when $t$ is small). Together with the discussions in Section~\ref{sec:decoding of SCLD}, we conclude that SCLDs have not only much more efficient traceability than SCs but also much larger code rate than FPCs.

\section{Bounds for binary $\bar{t}$-SCLDs with constant list size}
\label{sec: binary SCLD constant list size}

In this section we consider the existence of binary SCLDs with constant list size. In particular, we establish a lower bound on the 
largest asymptotic code rate of SCLDs with constant list size. 
In contrast to the previous Section~\ref{sec: binary SCLD} considering very large list size, we need to execute more careful analyses for the case with constant list size here.   
The following Markov's inequality and Hoeffding's inequality will be exploited.  

\begin{lemma}[e.g. \cite{chernoff1952measure}]
\label{lemma-markov-hoffeding}
Let $Y$ be a random variables such that $Y\ge 0$, and $Z_1,\ldots,Z_N$ be independent random variables such that $a_i\le Z_i\le b_i$ for all $i\in [N]$. 
Let $S_N= Z_1+\cdots+Z_N$. For any $\gamma >0$, we have 
\begin{enumerate}
    \item 
    $\text{Pr}(Y\ge \gamma) \le \frac{\E[Y]}{\gamma}$ \quad (Markov's inequality); 
    \item 
    $\text{Pr}(|S_N - \E[S_N]|\ge \gamma) \le 2\cdot \text{exp}\Big(-\frac{2\gamma^2}{\sum_{i=1}^N (b_i-a_i)^2}\Big)$\quad (Hoeffding's inequality),
\end{enumerate}
where $\E[Y]$ is the expectation of $Y$. 
\end{lemma}

\begin{theorem}\label{thm-scld-constant-L}
Let $t\ge 2$ be a positive integer. Then 
\begin{align}
    R_{\rm SCLD}(\bar{t}; L)\geq \max\limits_{p\in(0, 1/2]}\min\big\{\underline{R}_{\rm SC}(t, p),
    \underline{R}_{\rm HLD}(t, p, L)\big\},
    \end{align}
where $\underline{R}_{\rm SC}(t, p)$ is defined in~\eqref{eq:rate of sep codes} and
\begin{align}
    \label{def-R_hld-1}
    \underline{R}_{\rm HLD}(t, p, L)&\triangleq h(p)+\frac{B(t, p, L)}{L}\\
    B(t, p, L)&=p\log_2\left(\frac{q_1(t, L, 1-z)}{q_1(t, L, 1-z)+q_2(t, L, 1-z)}\right)
    +(1-p)\log_2\left(\frac{q_1(t, L, z)}{q_1(t, L, z)+q_2(t, L, z)}\right)\\
    q_1(t, L, z)&=z^t(z-z^t)^{L-t+1}\\
    q_2(t, L, z)&=(z-z^t)(1-z^t-(1-z)^t)^{L-t+1}
\end{align}
and $z\in(0, 1)$ is a unique root of the equation
\begin{align}
\label{def-R_hld-5}
p(q_1(t, L, z)+q_2(t, L, z))
=(1-p)(q_1(t, L, 1-z)+q_2(t, L, 1-z)).
\end{align}
\end{theorem}

\begin{IEEEproof}
Let $\mathcal{C}=\{\c_1,\c_2,\ldots,\c_M\}$ be a collection of $M=2^{nR}$ binary vectors of length $n$, in which each coordinate $\c_i(j)$ is chosen from $\{0,1\}$ independently at random and equals $1$ with probability $p$, where $p\in (0,1/2]$. 
We now aim to remove some vectors, which violate the definition of SCLD, from $\C$.  
To that end, we define bad items as follows. 
\begin{itemize}
    \item 
    Call a pair of distinct index sets $(I_1,I_2)$, where $I_1\subseteq [M]$, $I_2\subseteq [M]$, $|I_1|=t_1$, $|I_2|=t_2$, $|I_1\cap I_2|=m$, $1 \le t_1, t_2\le t$, a \textit{bad $(t_1, t_2, m)$-pair} if $\text{desc}(\{\c_i: i \in I_1\})=\text{desc}(\{\c_i: i \in I_2\})$. Denote the number of bad $(t_1, t_2, m)$-pairs for all $(t_1, t_2, m)\in \Xi$ as $X_1$, where $\Xi$ is defined as \eqref{def-Xi}. 
    \item 
    Call a vector $\c_i$ \textit{bad} if its weight is not equal to $\lfloor p(n+1)\rfloor$; otherwise, call it \textit{good}. Denote the set of all good vectors $\c_i$ in $\C$ as $\C_0$. Clearly $\C_0\subseteq \C$. Denote the number of good vectors in $\C$ as $X_2$ and $X_2=|\C_0|$. 
    \item   
    Call a pair of distinct sets $(\C_1, \C_2)$, where $\C_1, \C_2 \subseteq \C_0$, $|\C_1|=t$, $|\C_2|=L+1$, \textit{bad} if $\text{desc}(\C_1)$ covers all codewords from $\C_2$. Denote the number of bad pairs of sets in $\C_0$ as $X_3$. 
\end{itemize}
Now we would like to prove the following three claims.
\begin{claim}\label{claim-1}
$\text{Pr}\Big(X_1< \frac{M}{16(n+1)}\Big)> \frac{3}{4}$. 
\end{claim}
In fact, the code $\C$ is the same random code as in the proof of Theorem~\ref{thm-lower-scld}, where we estimated the mathematical expectation of the number of bad $(t_1, t_2, m)$-pairs as $\E[X_1]<c M/(n+1)$ in the case that $R<\underline{R}_{\rm SC}(t, p)$ for an arbitrary constant $c$ and large enough $n$. Here we take $c=1/64$. Then we obtain  
\begin{align*}
    \text{Pr}\Big(X_1< \frac{M}{16(n+1)}\Big) = 1-\text{Pr}\Big(X_1\ge \frac{M}{16(n+1)}\Big) 
    \ge 1- \frac{\E[X_1]}{\frac{M}{16(n+1)}} > 1- \frac{\frac{M}{64(n+1)}}{\frac{M}{16(n+1)}} = \frac{3}{4}
\end{align*}
where the first inequality follows from the Markov's inequality in Lemma~\ref{lemma-markov-hoffeding}. This proves Claim~\ref{claim-1}. 

\begin{claim}\label{claim-2}
$\text{Pr}\Big(X_2\ge \frac{M}{2(n+1)}\Big)= 1-o(1)$ as $n\to \infty$. 
\end{claim}

In fact, recall that, in the proof of Theorem~\ref{thm-lower-scld}, we showed that the probability that a vector is good is at least $1/(n+1)$. It implies that the expectation of the number of good vectors $\E[X_2]\ge M/(n+1)$. Thus 
\begin{align}\label{ineq-from-hoff}
    \text{Pr}\Big(X_2< \frac{M}{2(n+1)}\Big) 
    \le \text{Pr}\Big(|X_2-\E[X_2]|\ge  \frac{M}{2(n+1)}\Big) 
    \le  2\exp\left(-\frac{M}{2(n+1)^2}\right)
\end{align}
where the second inequality follows from the Hoeffding’s inequality in Lemma~\ref{lemma-markov-hoffeding}. Now we have 
\begin{align*}
    \text{Pr}\Big(X_2\ge \frac{M}{2(n+1)}\Big)
    = 1- \text{Pr}\Big(X_2< \frac{M}{2(n+1)}\Big) 
    \ge 1- 2\exp\left(-\frac{M}{2(n+1)^2}\right)
    = 1-o(1)\quad (n\to \infty)
\end{align*}
where the inequality follows from \eqref{ineq-from-hoff}. 
This proves Claim~\ref{claim-2}.

\begin{claim}\label{claim-3}
$\text{Pr}\Big(X_3< \frac{3X_2}{4}\Big)\ge \frac{1}{3}$. 
\end{claim}

In fact, notice that the code $\C_0$ consisting of all good vectors from $\C$ can be seen as a random code with a fixed weight $\lfloor p(n+1)\rfloor$ and cardinality $|\C_0|=X_2$, whose distribution is the same as in the case when every codeword is taken independently and equiprobably from the set of all vectors of weight $\lfloor p(n+1)\rfloor$. 
This fact coincides with the assumption in the proof of~\cite[Theorem 3]{shchukin2016list}, where a fixed weight ensemble was considered as well. 
It was shown in \cite[Theorem 3]{shchukin2016list} that if the code rate $R_{\C_0}=\frac{1}{n}\log_2 (X_2)$ of $\C_0$  is smaller than $\underline{R}_{\rm HLD}(t, p, L)$, where $\underline{R}_{\rm HLD}(t, p, L)$ is defined from \eqref{def-R_hld-1}-\eqref{def-R_hld-5}, then the mathematical expectation of the number $X_3$ of bad pairs of sets satisfies $\E[X_3]\le X_2/2$ for large enough $n$. 
Thus we have 
\begin{align*}
    \text{Pr}\Big(X_3< \frac{3X_2}{4}\Big)
    = 
    1-\text{Pr}\Big(X_3\ge \frac{3X_2}{4}\Big)
    \ge 1- \frac{\E[X_3]}{\frac{3X_2}{4}}
    \ge 1- \frac{\frac{X_2}{2}}{\frac{3X_2}{4}}
    = \frac{1}{3} 
\end{align*}
where the first inequality follows from the Markov’s inequality in Lemma~\ref{lemma-markov-hoffeding}.
This proves Claim \ref{claim-3}. 

Based on Claims \ref{claim-1}-\ref{claim-3}, we conclude that 
if $R <  \min\big\{\underline{R}_{\rm SC}(t, p),
    \underline{R}_{\rm HLD}(t, p, L)\big\}$,
with positive probability all three conditions below could be satisfied: 
\begin{align*}
   X_1< \frac{M}{16(n+1)} \ \ \text{and} \ \ 
   X_2\ge \frac{M}{2(n+1)} \ \ \text{and}\ \ 
   X_3< \frac{3X_2}{4}.    
\end{align*}
By this, from the code $\C_0$ we delete one vector from each bad pair of sets and one vector from each bad $(t_1, t_2, m)$-pair for all $(t_1, t_2, m)\in \Xi$. It is easily verified that the resulting code is a 
$\bar{t}$-SCLD$\big(n,M_0,2;L)$ with code size 
$$M_0\ge X_2-X_1-X_3> 
\frac{X_2}{4} - X_1
>
\frac{M}{8(n+1)} - \frac{M}{16(n+1)} = 
\frac{M}{16(n+1)}$$
and code rate $\frac{1}{n} \log_2M_0 \ge R+o(1)$ as $n\to \infty$.
This completes the proof of Theorem \ref{thm-scld-constant-L}. 
\end{IEEEproof}

\vskip 0.1cm 

Table \ref{table:SCLD-constant-L} illustrates the numerical calculation results for $t=2, 3$ from Theorem~\ref{thm-scld-constant-L}. 
It is worth noting from the case $t=3$ that even the list size is not sufficiently large, the code rate of binary SCLDs is very good already, in comparison with Table~\ref{table:numerical_results} concerning asymptotically large list size. 
In the next Section \ref{sec: q-ary SCLD}, 
we will show this interesting phenomenon is more manifest for $q$-ary SCLDs with sufficiently large alphabet size $q$. 

\begin{table}[h]\centering
\caption{Lower bounds for binary
SCLDs with constant list size via Theorem \ref{thm-scld-constant-L}}
\label{table:SCLD-constant-L}
\begin{tabular}{|c|c|c|c|c|c|c|} 
\hline
 $(t,L)$&  $(2,3)$ & $(2,4)$ & $(2,5)$ & $(2,6)$ & $(2,7)$
 \\
\hline 
$R_{\rm SCLD}(\bar{t},L)\ge $ & 
$0.245655$ &
$0.263492$ &
$0.274428$ &
$0.281927$ &
$0.287402$ 
\\
\hline
 $(t,L)$&  $(3,4)$ & $(3,5)$ & $(3,6)$ & $(3,7)$ & $(3,8)$
\\
\hline  
$R_{\rm SCLD}(\bar{t},L)\ge $ & 
$0.115118$ &
$0.126598$ &
$0.129504$ &
$0.130385$ &
$0.130601$ 
\\
\hline
\end{tabular}
\end{table}

\section{Bounds for $q$-ary SCLDs}
\label{sec: q-ary SCLD}

In this section we provide lower bounds for the largest possible code rate of $q$-ary SCLDs with large $q$. Interestingly, our bound for $q$-ary SCLDs matches the (almost optimal) bound of $q$-ary SCs, which implies that the established $q$-ary SCLDs are almost optimal as well. 
Precisely, we prove the following theorem. 

\begin{theorem}\label{thm-lower-qnary}
Let $t,n,L$ be positive integers. Then 
\begin{equation*}
	R_{\rm SCLD}(\bar{t}, n; L)\ge 
	\begin{cases} 
	2/3 & \text{if $t=2$ and $L\ge 3$,}\\ 
	1/(t-1) & \text{if $t>2$ and $L\ge t+1$}. 
	\end{cases} 
\end{equation*}
\end{theorem}

\begin{IEEEproof}
	Let $\mathcal{C}=\{\c_1,\c_2,\ldots,\c_M\}$ be a collection of $M=\varepsilon(n, t) q^{nR_{\delta}}$ $q$-ary vectors of length $n$, in which each coordinate $\c_i(j)$ is chosen independently at random and equal to $k$ with probability $1/q$ for each $k\in Q$,  
	$\varepsilon(n, t)$ is a sufficiently small positive constant less than $1$ depending only on $n$ and $t$, 
	and 
	\begin{equation}
	R_{\delta}= 
	\begin{cases} 
	2/3-\delta & \text{if $t=2$, }\\ 
	1/(t-1)-\delta & \text{if $t>2$} 
	\end{cases} 
	\end{equation}
for an arbitrary small $\delta >0$.
Next we discuss the two requirements of SCLD in Definition~\ref{def-scld}.

(I) Consider the first requirement of the separable property. We use the following result from~\cite{B15}. 
\begin{claim}[\hspace{1sp}{\cite[Theorems 4 and 5]{B15}}]
	    For arbitrary small $\delta>0$ and sufficiently small $0 < \varepsilon(n, t)< 1$  there exists a random set $\C'\subseteq \C$ 
	    such that 
	    \begin{itemize}
	        \item for all distinct $\mathcal{C}_1,\mathcal{C}_2\subseteq \mathcal{C}'$ with $|\mathcal{C}_1|\le t$ and $|\mathcal{C}_2|\le t$, it holds that  $\text{desc}(\mathcal{C}_1)\ne \text{desc}(\mathcal{C}_2)$; and
	        \item the expectation of $\vert\C'\vert$ is at least $\kappa q^{nR_{\delta}}$, where $\kappa$ is any positive constant less than ${\varepsilon}-{\varepsilon}^2{\kappa}'$, and ${\kappa}'$ is a constant depending only on $n$ and $t$.
	    \end{itemize}
\end{claim}
In other words, it is possible to remove a small number of vectors (in $\C\setminus \C'$) from $\C$ such that the set of remained vectors $\C'\subseteq \C$ meets the first requirement of SCLD in Definition~\ref{def-scld}. 

(II) We now consider the second requirement of SCLD in Definition~\ref{def-scld}. 
   For $\C_0\subseteq \C$, $|\C_0|= t$, we call $\C_0$ a \textit{bad set} if 
   $$|\text{Res}_{\C}(\text{desc}(\C_0))|>L.$$
   Let $E$ denote the expectation of the number of bad sets in $\C$. We are going to estimate $E$. 
   
	Consider a fixed evidence vector $\mathbf{d}$ produced by a coalition of $t$ colluders holding fingerprints in $\mathcal{C}$. 
	The probability of that $\mathbf{d}$ covers more than $L$ vectors in $\C$ is upper bounded by 
	\begin{align*}
	\binom{M}{L+1-t}(t/q)^{n(L+1-t)}
	&\le 
	\frac{M^{L+1-t}}{(L+1-t)!} q^{n(\log_qt-1)(L+1-t)}\\		
	&\le\frac{M^{L+1-t}}{(L+1-t)!}\Big(\frac{M}{\varepsilon}\Big)^{R_{\delta}^{-1}(\log_qt-1)(L+1-t)}\\
	&\leq
	\Big(\frac{M}{\varepsilon}\Big)^{(L+1-t)(1+R_{\delta}^{-1}(\log_qt-1))}/(L+1-t)!.
\end{align*}
By the linearity of expectation, 
the expectation $E$ of the number of bad sets is at most
$$
E<M^t\Big(\frac{M}{{\varepsilon}}\Big)^{(L+1-t)(1+R_{\delta}^{-1}(\log_qt-1))}/(L+1-t)!
$$
Specifically, we have the following for an arbitrary small $\delta>0$: 
\begin{itemize}
    \item For $t=2$, $R_{\delta}=2/3-\delta$, $L=3$, and $q>2^{\frac{2}{3\delta}}$, we obtain  
    $$E<\frac{1}{2}\Big(\frac{M}{\varepsilon}\Big)^{4+2\frac{\log_q2-1}{2/3-\delta}}=o(M)=o(q^{nR_{\delta}}).$$
    \item 
    
For $t>2$, $R_{\delta}=\frac{1}{t-1}-\delta$, $L= t+1$, and $q>t^{\frac{2t-2}{t-3+(t^2-1)\delta}}$, we have
\begin{align*}
    E&<\frac{1}{2}\Big(\frac{M}{\varepsilon}\Big)^{t+2+2\frac{\log_qt-1}{1/(t-1)-\delta}}
    =o(M)=o(q^{nR_{\delta}}).
\end{align*}
\end{itemize}

Based on (I) and (II), we remove the vectors in $\C\setminus\C'$ and also remove a vector from each bad set in $\C$. It is easily verified the obtained code is a $\bar{t}$-SCLD, and the expectation of code size is at least $\kappa q^{nR_{\delta}}(1-o(1))$. Therefore, together with Lemma~\ref{lemma:code-relations}, we conclude that there exists a $\bar{t}$-SCLD$(n, M', q; L)$ with $L\ge t+1$ and $M'\ge\kappa' q^{nR_{\delta}}$, where $\kappa'$ is a constant depending only on $n$ and $t$. 
In other words, we have shown that the code rate of $q$-ary $\bar{t}$-SCLD codes is at least $R_\delta$ for any $\delta>0$. Taking limit as $\delta\to 0$ we complete the proof.
\end{IEEEproof}
 
\vskip 0.05cm 
It is quite remarkable that the code rate of SCLDs in Theorem~\ref{thm-lower-qnary} achieves the best-known (and almost optimal) code rate of SCs~\cite{B15,GG}. 
Also we notice an intriguing property that the list size in $\bar{t}$-SCLDs just needs to be a little bit larger than $t$, i.e. $L\ge t+1$. Together with Section~\ref{sec:decoding of SCLD}, we conclude that SCLDs could have not only the same code rate as SCs for large $q$ but also much more efficient decoding than SCs.

\section{Constructions for SCLDs}
\label{sec:constructions}

In this section, we provide explicit constructions for SCLDs and discuss their corresponding identifying algorithms. 

\subsection{SCLDs from generalized packings}
\label{subsec: conclusion-GP}

Since the code length of a $\bar{t}$-SCLD corresponds to the number of orthonormal basis signals in the multimedia content, and the code size corresponds to the number of authorized users, it is thus desirable to construct $\bar{t}$-SCLDs with large code size while keeping their lengths short. 
In this subsection, we construct $\bar{2}$-SCLD$(2,M,q';3)$ from a combinatorial structure called \textit{generalized packings} \cite{colbourn}. 

\begin{definition}
Let $K$ be a subset of nonnegative integers, and let $v,b$ be two positive integers. A generalized $(v,b,K,1)$ packing is a set system $(X,{\cal B})$ where $X$ is a set of $v$ elements and ${\cal B}$ is a set of $b$ subsets of $X$ called blocks that satisfy
\begin{enumerate}
    \item[1)] $|B| \in K$ for any $B \in {\cal B}$; 
    \item[2)] every pair of distinct elements of $X$ occurs in at most one block of ${\cal B}$.
\end{enumerate}
\end{definition}

In a generalized $(v,b,K,1)$ packing, if $K = \{k\}$ for some $k>1$ and every pair of distinct elements of $X$ occurs in exactly one block, then it is usually called a \textit{balanced incomplete block design}, or briefly $(v,k,1)$-BIBD. A $(q^2+q+1,q+1,1)$-BIBD with $q \ge 2$ corresponds to a projective plane of order $q$. 

\begin{construction}
\label{cons:packing}
Let $(X,{\cal B})$ be a generalized $(v,v,K,1)$ packing with $X= \{0,1,\ldots,v-1\}$ and ${\cal B} = \{B_0, B_1, \ldots,$ $B_{v-1}\}$. 
Then ${\cal C} = C_0\cup C_1\cup \cdots\cup C_{v-1}$, with $C_i= \{(i,b): b \in B_i\}$ if $B_i \ne \emptyset$ and $C_i= \emptyset$ if $B_i = \emptyset$, is a $\bar{2}$-SCLD$(2,M,v;3)$ defined over $X$ with $M = |B_0| + |B_1| + \cdots + |B_{v-1}|$. 
\end{construction}

\begin{IEEEproof}
It is shown in \cite{CJM} that ${\cal C}$ is a $\bar{2}$-SC$(2,M,v)$ defined over $X$ with $M = |B_0| + |B_1| + \cdots + |B_{v-1}|$. To prove its list size is $3$, notice that $\text{desc}(\{(i,a),(j,b)\})$ cannot cover any additional codeword if $i=j$, and covers at most one codeword $(i,b)$ or $(j,a)$ if $i \ne j$, but not both. Otherwise, $\text{desc}(\{(i,a),(j,b)\})$ would be equal to  $\text{desc}(\{(i,b),(j,a)\})$, a contradiction to the fact that ${\cal C}$ is a $\bar{2}$-SC$(2,M,v)$. This completes the proof.
\end{IEEEproof}

As an immediate consequence of Construction \ref{cons:packing}, we have the following. 

\begin{corollary}
\label{coro:plane}
For any prime power $q$, there exist 
\begin{enumerate}
    \item[1)] a $\bar{2}$-SCLD$(2,(q+1)(q^2+q+1), q^2+q+1;3)$; 
    \item[2)] a $\bar{2}$-SCLD$(2,q^3+2q^2, q^2+q;3)$. 
\end{enumerate}
\end{corollary}

\begin{IEEEproof}
Claim 1) follows from the well-known fact that there exists a projective plane of order $q$ if $q$ is a prime power~\cite{colbourn}. Delete one block from a projective plane and an element  from this block, we obtain a generalized  $(q^2+q,q^2+q,\{q,q+1\},1)$ packing with $q$ blocks of size $q$ and $q^2$ blocks of size $q+1$. This gives a $\bar{2}$-SCLD$(2,q^3+2q^2, q^2+q;3)$, as desired. 
\end{IEEEproof}

According to Lemma \ref{lemma:code-relations}, the largest code size of $\bar{t}$-SCLDs can be upper bounded by that of $\bar{t}$-SCs. By \cite{CJM}, we conclude that the above two $\bar{2}$-SCLDs of length $2$ have the largest possible code size, respectively, given the alphabet size.  

\subsection{A concatenated construction}
In this subsection, we provide a concatenated construction which allows to derive SCLDs with small alphabet size from SCLDs with large alphabet size (e.g. the constructions in Section \ref{subsec: conclusion-GP}). 

\begin{construction}
\label{cons:concatenation}
Let $2\le q \le q'$ be two integers. 
Let ${\mathcal A}$ be a $t$-FPC$(n_1,q',q)$ over the alphabet $Q$ and ${\mathcal B}$ be a $\bar{t}$-SCLD$(n_2,M,q';L)$ over the alphabet $Q'$. Define a bijection $\phi:Q' \to {\mathcal A}$. Let ${\mathcal C}$ be the code defined by 
\begin{align*}
    \Phi: {\mathcal B} &\to {\mathcal C}\\
    {\bf b} = ({\bf b}(1), \ldots, {\bf b}(n_1)) &\mapsto \Phi({\bf b}) = (\phi({\bf b}(1)), \ldots, \phi({\bf b}(n_1))).
\end{align*}
Then ${\mathcal C}$ is a $\bar{t}$-SCLD$(n_1n_2,M,q;L)$ over $Q$.
\end{construction}

\begin{IEEEproof}
It is obvious that ${\mathcal C}$ is an $(n_1n_2,M,q)$ code over $Q$. Since any $t$-FPC is a $\bar{t}$-SC, by  \cite[Lemma 5.4]{CM}, we know that ${\mathcal C}$ is a $\bar{t}$-SC$(n_1n_2,M,q)$. It suffices to show that ${\mathcal C}$ is also a $t$-HLD$(n_1n_2,M,q; L)$. Notice that ${\mathcal A}$ is a $t$-FPC$(n_1,q',q)$ so that no codeword outside a coalition can be covered by the descendant code of the coalition, we know that any coalition $\{\Phi({\bf b}_1), \ldots, \Phi({\bf b}_s)\}$, $s \le t$, in ${\mathcal C}$ corresponds exactly to a coalition $\{{\bf b}_1, \ldots, {\bf b}_s\}$ in ${\mathcal B}$. This implies that there are at most $L$ codewords in ${\mathcal C}$ covered by $\text{desc}(\{\Phi({\bf b}_1), \ldots, \Phi({\bf b}_s)\})$, i.e., ${\mathcal C}$ is a $t$-HLD$(n_1n_2,M,q; L)$. According to Lemma \ref{lemma:code-relations}, the conclusion follows. 
\end{IEEEproof}

\vskip 0.1cm
In order to use Construction \ref{cons:concatenation} to derive SCLDs, we recall two known constructions of frameproof codes as follows.

\begin{construction}[\hspace{1sp}\cite{B03}]
\label{cons:B1}
Let $l\ge 2$ and $t\ge 2$ be two integers, and $q \ge l$ be a prime power. Let ${\alpha}_1, {\alpha}_2, \ldots, {\alpha}_l \in \text{GF}(q)$ be distinct. Define a code ${\mathcal C}$ over $\text{GF}(q)$ by 
$$ {\mathcal C} = \{(f({\alpha}_1), f({\alpha}_2), \ldots,f({\alpha}_l)): f \in \text{GF}(q)[x] {\rm \ and \ deg}(f) < \lceil l/t \rceil\}.$$
Then ${\mathcal C}$ is a $t$-FPC$(l,q^{\lceil l/t \rceil}, q)$. 
\end{construction}

\begin{construction}[\hspace{1sp}\cite{B03}]
\label{cons:B2}
Let $l \ge 4$ be an even integer. Let $m \ge l+1$ be a prime power and $q=m^2+1$. Let ${\beta}_0, {\beta}_1, {\alpha}_1, {\alpha}_2, \ldots, {\alpha}_{l-1} \in \text{GF}(m)$ be distinct. Define  
\begin{align*}
{\mathcal C}_1 &= \Big\{(\infty,(f({\alpha}_1),g({\alpha}_1)),  
\ldots,(f({\alpha}_{l-1}),g({\alpha}_{l-1})): f,g \in \text{GF}(m)[x], {\rm \ deg}(f) = \frac{l}{2}-1, {\rm deg} (g) \le  \frac{l}{2}-1\Big\},\\
{\mathcal C}_2 &= \Big\{\big((t({\beta}_0),t({\beta}_1)),(s({\alpha}_1),t({\alpha}_1)),
\ldots,(s({\alpha}_{l-1}),t({\alpha}_{l-1}))\big): s,t \in \text{GF}(m)[x], {\rm \ deg}(s) \le \frac{l}{2}-2, {\rm deg}(t) \le  \frac{l}{2}\Big\}.
\end{align*}
Then the code ${\mathcal C} = {\mathcal C}_1 \cup {\mathcal C}_2$ is a $2$-FPC$(l,2(q-1)^{l/2}(1-1/(2\sqrt{q-1}),q)$ defined over $F = \{\infty\} \cup (\text{GF}(m))^2$.
\end{construction}

Apply Construction~\ref{cons:concatenation} with Corollary~\ref{coro:plane} and Constructions~\ref{cons:B1}, \ref{cons:B2}, we immediately obtain the followings.  

\begin{corollary}
\label{coro:scld}
Let $n$ be a prime power.
\begin{enumerate}
    \item[1)] For any integer $l \ge 2$, there exists a $\bar{2}$-SCLD$(2l,(n+1)(n^2+n+1),q;3)$ for any prime power $q$ such that $q \ge l$ and $q^{\lceil l/2 \rceil} \ge n^2+n+1$. 
    \item[2)] For any even $l \ge 4$, there exists a $\bar{2}$-SCLD$(2l,(n+1)(n^2+n+1),q;3)$ for any $q=m^2+1$ where $m\ge l+1$ is a prime power such that $2(q-1)^{l/2}(1-1/(2\sqrt{q-1})) \ge n^2+n+1$.
    \item[3)] For any integer $l \ge 2$, there exists a $\bar{2}$-SCLD$(2l,n^3+2n^2,q;3)$ for any prime power $q$ such that $q \ge l$ and $q^{\lceil l/2 \rceil} \ge n^2+n$. 
    \item[4)] For any even $l \ge 4$, there exists a $\bar{2}$-SCLD$(2l,n^3+2n^2,q;3)$ for any $q=m^2+1$ where $m\ge l+1$ is a prime power such that $2(q-1)^{l/2}(1-1/(2\sqrt{q-1})) \ge n^2+n$.
\end{enumerate}
\end{corollary}

\subsection{An algebraic construction with efficient decoding}

In this subsection we show an algebraic construction for binary $\bar{2}$-SCLDs with both of a high code rate and an efficient identifying algorithm in time $O(\polylog(M))$.  
The following construction 
is originally from \cite{L69}. 

\begin{construction}[\hspace{1sp}\cite{L69}]
\label{construction:2-SCLD}
Let $l\ge 2$ be an integer. For each element $\mathbf{x}\in \GF(2^l)$, $\mathbf{x}$ can be represented as a binary vector of length $l$ \cite{LN}. 
Define a binary code $\C$ as 
\begin{align*}
    \C = \{(\mathbf{x},\mathbf{x^3}):\, \mathbf{x}\in \GF(2^l)\}.
\end{align*}
Then $\C$ is a $\bar{2}$-SCLD$(2l, 2^l,2;2^l)$.
\end{construction} 

It is readily seen that the above $\bar{2}$-SCLDs with list size $L=M=2^l$ has code rate $1/2$, which is larger than the code rate obtained from the random coding method in Theorem~\ref{thm-lower-scld}.  
Notably, 
we find that the $\bar{2}$-SCLDs via Construction \ref{construction:2-SCLD} could perform identification very efficiently as follows.

\begin{theorem}
\label{thm:fast_decoding_alg}
For a $\bar{2}$-SCLD$(n,M,2;M)$ derived from Construction \ref{construction:2-SCLD}, there exists an identifying algorithm with time complexity 
$O(\poly(n))=O(\polylog(M))$.
\end{theorem}

\begin{algorithm}[t]
\caption{A solution for the quadratic equation \eqref{eq:quadratic_z} over $\GF(2^l)$ \cite{Chen82}}\label{Alg:find_solution_quadratic}
\textbf{Input:} 
a primitive element $\omega$ of $\GF(2^l)$ such that the \textit{trace} of $\omega$ equals to $1$, i.e. $\sum\limits_{i=0}^{l-1}\omega^{2^i}=1$;\\
\text{\quad \quad \ \ \,}
the constant term $k=\mathbf{1} + \mathbf{v}/\mathbf{u^3}\in \GF(2^l)$
\vskip 0.1cm
\begin{algorithmic}[1]
\If {$l\pmod{2}== 1$}
\Comment{Case 1}
\vskip 0.05cm
\State 
$\mathbf{z} = \sum\limits_{i=0}^{(l-1)/2} k^{2^{2i}}$
\Else
\State 
$T(k)=\sum\limits_{i=0}^{(l-2)/2} k^{2^{2i}}$
\EndIf 
\If {$l\pmod{4}== 2$ and $T(k)==0$}
\Comment{Case 2}
\vskip 0.1cm
\State 
$\mathbf{z} = \sum\limits_{i=0}^{(l-6)/4} (k+k^2)^{2^{2+4i}}$
\ElsIf{$l\pmod{4}== 2$ and $T(k)==1$}
\Comment{Case 3}
\vskip 0.1cm
\State \qquad 
$\mathbf{z} =
\omega^{(2^l-1)/3} + \sum\limits_{i=0}^{(l-6)/4} (k+k^2)^{2^{2+4i}}$
\vskip 0.1cm
\ElsIf{$l\pmod{4}== 0$ and $T(k)==1$}
\Comment{Case 4}
\vskip 0.1cm
\State \qquad 
$S=\sum\limits_{j=1}^{l/4-1}\sum\limits_{i=j}^{l/4-1} k^{2^{2i-1+l/2} + 2^{2j-2}}$
\State \qquad 
$\mathbf{z}= S + S^2 +k^{2^{l-1}}\bigg(1+\sum\limits_{i=0}^{l/4-1} k^{2^{2i+l/2}} \bigg)$
\vskip 0.1cm
\ElsIf{$l\pmod{4}== 0$ and $T(k)==0$}
\Comment{Case 5}
\vskip 0.1cm
\State \qquad 
$S_1=\sum\limits_{j=1}^{l/4-1}\sum\limits_{i=j}^{l/4-1} (\omega + \omega^2+k)^{2^{2i-1+l/2} + 2^{2j-2}}$
\State \qquad 
$\mathbf{z}= \omega + S_1 + S_1^2 +(\omega + \omega^2+k)^{2^{l-1}}\bigg(1+\sum\limits_{i=0}^{l/4-1} (\omega + \omega^2+k)^{2^{2i+l/2}} \bigg)$
\vskip 0.1cm
\EndIf 
\end{algorithmic}
\textbf{Output:} a solution $\mathbf{z}\in \GF(2^l)$ for equation \eqref{eq:quadratic_z}
\end{algorithm} 

\begin{IEEEproof}
Let $\C = \{(\mathbf{x},\mathbf{x^3}):\, \mathbf{x}\in \GF(2^l)\}$ be a $\bar{2}$-SCLD$(n,M,2;M)$ derived from Construction \ref{construction:2-SCLD} with $n=2l$ and $M=2^l$. For any coalition of size two, say $(\mathbf{x},\mathbf{x^3})$ and $(\mathbf{y},\mathbf{y^3})$ with $\mathbf{x}\neq  \mathbf{y}\in \GF(2^l)$, their evidence vector $\d$ implies $(\mathbf{u},\mathbf{v})$, $\mathbf{u},\mathbf{v}\in \GF(2^l)$, such that 
\begin{align}
\label{eq:decoding_two_eqs}
    \mathbf{x} + \mathbf{y} = \mathbf{u}
    \quad 
    \text{and}
    \quad 
    \mathbf{x^3} + \mathbf{y^3} = \mathbf{v},  
\end{align}
where $\mathbf{u}(i) = 1$ if $\mathbf{d}(i)=\{0,1\}$, $\mathbf{u}(i) = 0$ if $\mathbf{d}(i)=\{0\}$ or $\{1\}$, 
and $\mathbf{v}(i) = 1$ if $\mathbf{d}(l+i)=\{0,1\}$, $\mathbf{v}(i) = 0$ if $\mathbf{d}(l+i)=\{0\}$ or $\{1\}$ for $1 \le i \le l$. 
Clearly, $\mathbf{u}\neq 0$. 
The identifying aims to find the solutions $\mathbf{x}$ and $\mathbf{y}$ using $\mathbf{u}, \mathbf{v}$ in \eqref{eq:decoding_two_eqs}. 
To that end, we first note that \eqref{eq:decoding_two_eqs} can be transferred to 
    $\mathbf{x^2} + \mathbf{u}\cdot \mathbf{x} +  \mathbf{u^2} + \mathbf{v}/\mathbf{u} = \mathbf{0}.$ 
Let $\mathbf{z}=\mathbf{x}/\mathbf{u}$. Then we have 
\begin{align}
\label{eq:quadratic_z}
    \mathbf{z^2} + \mathbf{z} +  \mathbf{1} + \mathbf{v}/\mathbf{u^3} = 0.
\end{align}
Next we show that one solution of \eqref{eq:quadratic_z} in $\GF(2^l)$ could be determined by means of Algorithm \ref{Alg:find_solution_quadratic}, in which we use a result 
that $\GF(2^l)$ contains a primitive element of $trace$ equal to $1$ from \cite[Theorem 2]{M82} as well.   

Indeed, Algorithm \ref{Alg:find_solution_quadratic} can determine a solution $\mathbf{z}$ of quadratic equation \eqref{eq:quadratic_z} in time $O(\poly(l))$. 
Together with the relation $\mathbf{x}=\mathbf{u}\cdot \mathbf{z}$ and the symmetry of $\mathbf{x}$ and $\mathbf{y}$, the solution of \eqref{eq:decoding_two_eqs} can be determined. 
In other words, 
the decoding of $\C$ (based on Algorithm \ref{Alg:find_solution_quadratic}) can be done in time 
$O(\poly(l))=O(\poly(n))=O(\polylog(M))$, as required. 
\end{IEEEproof}

\section{Two-stage dynamic traitor tracing}
\label{sec:dynamic_traitor_tracing}

In this section we discuss the two-stage traitor tracing for the dynamic scenario. In particular, we establish a two-stage dynamic traitor tracing framework based on HLDs and SCLDs. It is shown that it could provide not only more efficient decoding algorithms but also much larger code rate (i.e. accommodate more users) than SCLDs in the static scenario.

\begin{theorem}
\label{thm:dynamic}
Let $t$ be a positive integer and the number of traitors be no more than $t$. 
There exists a two-stage dynamic traitor tracing scheme accommodating 
$M=\Omega(2^{nR_{\rm TDTT}(\alpha,\beta,t)})$
users and with complete traceability in time $O(nM^{\max\{1,\alpha\beta t\}})$, where 
$R_{\rm TDTT}(\alpha,\beta,t)= \frac{1}{2} R^{(\alpha)}_{\rm HLD}(t)$ and $0\le \alpha, \beta \le 1$ such that 
$\alpha\cdot R^{(\alpha)}_{\rm HLD}(t) \le  R^{(\beta)}_{\rm SCLD}(\bar{t})$. 
\end{theorem}
 
\begin{IEEEproof}
Suppose the total number of authorized users is $M$ and the total number of traitors is at most $t$, where $t\le M$. 
We build a two-stage traitor tracing scheme as in Algorithm \ref{Alg:Dynamic}.

\begin{algorithm}
\caption{Two-stage dynamic traitor tracing algorithm}\label{Alg:Dynamic}\label{Alg2}
\textbf{Input:} $t$-HLD code $\mathcal{C}=\{\c_1, \ldots,\c_M\}$ with list size 
$L^{(1)}=M^{\alpha}$; the evidence vector 
$\mathbf{d}^{(1)}\in \mathcal{P}(Q)^n$
\Comment{Stage 1}

\begin{algorithmic}[1]
\State $\mathcal{T} = \emptyset $, $\mathcal{W} = \emptyset $. \Comment{Initialize the candidate sets}
\For{each $j \in [M]$} 
\If {$\mathbf{c}_j\preceq \mathbf{{d}}^{(1)}$}
\State $\mathcal{W} = \mathcal{W}\cup \{j\} $;
\EndIf
\EndFor
\end{algorithmic}
\textbf{Output:} the index set $\mathcal{W}$\\
\textbf{Input:} $\bar{t}$-SCLD code $\mathcal{C}=\{\c_1,\ldots,\c_{|\mathcal{W}|}\}$ with list size 
$L^{(2)}=|\mathcal{W}|^{\beta}$; the evidence vector  $\mathbf{d}^{(2)}\in \mathcal{P}(Q)^n$
\Comment{Stage 2}

\begin{algorithmic}[1]
\State execute the two steps as in Algorithm \ref{Alg:SCLD} for SCLD
\end{algorithmic}
\textbf{Output:} the set of all traitors $\mathcal{T}$
\end{algorithm}

In the first stage, 
we exploit a binary $t$-HLD$(n_0,M,2;L^{(1)})$ with the list size 
$L^{(1)}=M^{\alpha}$,
where $0\le \alpha\le 1$ is a constant and will be decided later. 
Correspondingly, the code size $M=2^{n_0R^{(\alpha)}_{\rm HLD}(t)}$, where $R^{(\alpha)}_{\rm HLD}(t)$ is defined as in \eqref{eq:def-R-hld-alpha}. 
According to the first half (i.e. Stage 1) of Algorithm~\ref{Alg:Dynamic}, given an evidence vector  $\mathbf{d}^{(1)}$ generated by the collusion attack,
a subset of users indexed by $\mathcal{W}$ could be identified in time $O(n_0M)$. 
By the definition of $t$-HLD, the size of $\mathcal{W}$ is upper bounded by 
$L^{(1)}=M^{\alpha}$
and all the traitors are in $\mathcal{W}$. In other words, in this stage we rule out some innocent users and narrow down the search space for the next stage such that all the traitors are in $\mathcal{W}$, which correspondingly conduces to further efficiently identify all exact traitors. 

In the second stage, we employ a binary $\bar{t}$-SCLD$(n_0,|\mathcal{W}|,2;L^{(2)})$ 
with the list size 
$L^{(2)}=|\mathcal{W}|^{\beta}$ 
where $0\le \beta\le 1$ is a constant and will be decided later. 
According to the latter part of Algorithm~\ref{Alg:Dynamic} (equivalently, Algorithm~\ref{Alg:SCLD}) and Theorem~\ref{thm:decoding}, given an evidence vector  $\mathbf{d}^{(2)}$ generated by the collusion attack, all the traitors could be traced back in time $O\big(\max\{n_0|\mathcal{W}|, n_0|\mathcal{W}|^{\beta t}\}\big)=O\big(n_0M^{\max\{\alpha,\alpha\beta t\}}\big)$. 

Let $n=2n_0$. Based on the foregoing, the total time cost of Algorithm~\ref{Alg2} is $O\big(nM^{\max\{1,\alpha\beta t\}}\big)$, and
the code rate (correspondingly, user capacity) of this two-stage dynamic traitor tracing is 
\begin{align}
\label{eq:def_R_TSDTT}
R_{\rm TDTT}(\alpha,\beta,t)= 
\limsup_{n\to\infty} \frac{\log_2 M}{n}
= \limsup_{n_0\to\infty}
\frac{\log_2 2^{n_0R^{(\alpha)}_{\rm HLD}(t)}}{2n_0}
= 
\frac{1}{2} R^{(\alpha)}_{\rm HLD}(t) .
\end{align}
Notice that in order to guarantee the existence of the corresponding $\bar{t}$-SCLD in the second stage, it is required that 
$$|\mathcal{W}|\le M^{\alpha} = 2^{\alpha\cdot n_0R^{(\alpha)}_{\rm HLD}(t)} \le 2^{n_0R^{(\beta)}_{\rm SCLD}(\bar{t})}.$$
In other words, $\alpha$ and $\beta$ need to satisfy 
$\alpha\cdot R^{(\alpha)}_{\rm HLD}(t) \le  R^{(\beta)}_{\rm SCLD}(\bar{t})$. 
Therefore the theorem follows. 
\end{IEEEproof}

\vskip 0.1cm
It is worth noting from Theorem \ref{thm:dynamic} that the choice of $\alpha,\beta$ plays an important role in finding a trade-off between the code rate (i.e. user capacity) and the tracing/identifying time complexity of two-stage dynamic traitor tracing. Roughly, if $\alpha$ and $\beta$ are larger (smaller), the code rate $R_{\rm TDTT}$ would be larger (smaller) while the identifying time complexity would be higher (lower).   
To see their performance precisely, we discuss two intriguing cases as below. 

\vskip 0.05cm
\textbf{Case 1.} \textit{To find the largest code rate $R_{\rm TDTT}$ without concerning the tracing time complexity.} 
To that end, we can set $\beta=1$ since $R_{\rm TDTT}(\alpha,\beta,t)$ is a non-decreasing function of $\beta\in [0,1]$. Accordingly, for each $t$, we aim to explore the optimum value of 
\begin{align}\label{eq-formulate-TDTT-case1}
    \max\ \frac{1}{2} R^{(\alpha)}_{\rm HLD}(t)  
    \text{\ \ subject to\ }\ \alpha\cdot R^{(\alpha)}_{\rm HLD}(t) \le  R^{(\beta)}_{\rm SCLD}(\bar{t}), \ \ \beta=1, \ \ 0<\alpha<1. 
\end{align}
The Table \ref{table:TDTT-case1} illustrates the numerical lower bounds for $R_{\rm TDTT}$ via \eqref{eq-formulate-TDTT-case1} together with Theorem~\ref{thm-lower-scld}, as well as its comparison with the code rates of SCLDs in the case when they have the same decoding complexity. 
It is easily seen that under the same complete traceability requirement, two-stage dynamic traitor tracing could have much larger code rate (i.e. accommodate much more users) than SCLDs.

\begin{table}[h]\centering
\caption{Lower bounds for TDTTs,
SCLDs when $\beta=1$}
\label{table:TDTT-case1}
\begin{tabular}{|c|c|c|c|c|c|c|} 
\hline
 $(\alpha, \beta, t)$&  $(0.40606,1,3)$ & $(0.29257,1,4)$ & $(0.21608,1,5)$ 
 \\
\hline 
$R_{\rm TDTT}(\alpha,\beta,t)$ & 
$0.16778$ &
$0.10224$ &
$0.07245$ 
\\
\hline  
$R^{(\alpha)}_{\rm SCLD}(\bar{t})$ & 
$0.13258$ &
$0.05783$ &
$0.03106$ 
\\
\hline
decoding cost& 
$O(nM^{1.21818})$  &
$O(nM^{1.17028})$ &
$O(nM^{1.08040})$ 
\\
\hline
\end{tabular}
\end{table}


\textbf{Case 2.} \textit{To find the largest code rate $R_{\rm TDTT}$ with tracing time complexity $O(nM)$.} 
From Theorem \ref{thm:dynamic}, it is seen that the time cost of two-stage traitor tracing is $O(nM^{\max\{1,\alpha\beta t\}})$, which takes the minimum value $O(nM)$ if $\alpha\beta \le 1/t$. Since $R_{\rm TDTT}(\alpha,\beta,t)$ is a non-decreasing function of $\alpha, \beta\in [0,1]$, we may consider the case when  $\alpha\beta = 1/t$ to explore the largest possible code rate. Accordingly, for each $t$, we aim to find the optimum value of 
\begin{align}\label{eq-formulate-TDTT-case2}
    \max\ \frac{1}{2} R^{(\alpha)}_{\rm HLD}(t)  
    \text{\ \ subject to\ }\ \alpha\cdot R^{(\alpha)}_{\rm HLD}(t) \le  R^{(\beta)}_{\rm SCLD}(\bar{t}), \ \ \alpha\beta=1/t, \ \ 0<\alpha, \beta <1. 
\end{align}
The Table \ref{table:TDTT-case2} illustrates the numerical lower bounds for $R_{\rm TDTT}$ from \eqref{eq-formulate-TDTT-case2} and Theorem~\ref{thm-lower-scld}. 
A comparison between SCLDs and two-stage dynamic traitor tracing under the same decoding cost $O(nM)$ in Table \ref{table:TDTT-case2} shows that two-stage dynamic traitor tracing could have much larger code rate (i.e. accommodate much more users) than SCLDs.

\begin{table}[h]\centering
\caption{Lower bounds for TDTTs,
SCLDs when $\alpha\beta=1/t$}
\label{table:TDTT-case2}
\begin{tabular}{|c|c|c|c|c|} 
\hline
 $(\alpha, \beta, t)$&  $(0.40406, 0.82496,3)$ & $(0.29107, 0.85890,4)$ &  $(0.21508, 0.92989,5)$ \\
\hline 
$R_{\rm TDTT}(\alpha,\beta,t)$ & 
$0.16722$ &
$0.10202$ &
$0.07236$ 
\\
\hline  
$R^{(1/t)}_{\rm SCLD}(\bar{t})$ & 
$0.13205$ &
$0.05770$ &
$0.03105$ \\
\hline
decoding cost& 
$O(nM)$ &
$O(nM)$  &
$O(nM)$ 
\\
\hline
\end{tabular}
\end{table}

\section{Conclusion}
\label{sec:conclusion}
In this paper we investigated combinatorial secure codes for traitor tracing. We initially integrated the list decoding idea directly into the practical model of traitor tracing with multimedia fingerprinting and proposed the notion of \textit{secure codes with list decoding} (SCLDs). It is shown that SCLDs can be seen as a unified concept in the sense that it could include many existing fingerprinting codes as special cases. We established efficient decoding/identifying algorithms and bounds on the largest possible code rate for SCLDs, which indicate that SCLDs could outperform the existing fingerprinting codes in terms of the decoding efficiency and/or the code rate. Furthermore, we proposed a two-stage traitor tracing framework for the dynamic scenario and showed that it has not only fast decoding but also much larger code rate than the static scenario. 
In the future work, it would be interesting to further improve the code rates and explore more explicit constructions for the binary SCLDs.

\section*{Acknowledgment}
Y. Gu would like to thank Prof. Tsuyoshi Takagi for inspirational suggestions on the study of traitor tracing with list decoding at an early stage, thank Prof. Minoru Kuribayashi for insightful comments on Section \ref{sec:dynamic_traitor_tracing}, and thank Prof. Shuichi Kawano for stimulating discussions.

\bibliographystyle{IEEEtranS}
\bibliography{refs}

\end{document}